\documentclass[
  aps,
  prl,
  reprint,
  superscriptaddress,
  nofootinbib,
  longbibliography
]{revtex4-2}
\usepackage{enumitem}
\usepackage{amsmath, amssymb}
\usepackage{graphicx}
\usepackage{xcolor}
\usepackage{textcomp}
\usepackage{hyperref}
\usepackage{orcidlink}
\usepackage{xspace}
\newlength{\mainfigureheight}
\newcommand{\Msun}{\ensuremath{\,\mathrm{M}_{\odot}}\xspace}

\definecolor{editcolor}{RGB}{0,120,100}

\hypersetup{
  colorlinks=true,
  linkcolor=blue,
  citecolor=blue,
  urlcolor=blue
}

\begin{document}

\title{Identity Crisis: Untangling Mass-Ordering Artifacts from Compact-Object Pairing in Gravitational-Wave Populations}

\author{Soumendra Kishore Roy\,\orcidlink{0000-0001-9295-5119}}
\email{sroy@cita.utoronto.ca}
\affiliation{Canadian Institute for Theoretical Astrophysics, University of Toronto, 60 St George St, Toronto, ON M5S 3H8, Canada}

\author{Asad Hussain\,\orcidlink{0000-0003-3491-5439}}
\affiliation{Center for Computational Astrophysics, Flatiron Institute, 162 Fifth Avenue, New York, NY 10010, USA}

\author{L. A. C. van Son\,\orcidlink{0000-0001-5484-4987}}
\affiliation{Radboud University, Faculty of Science, Huygensgebouw, Heyendaalseweg 135, 6525 AJ Nijmegen, The Netherlands}

\author{Maya Fishbach\,\orcidlink{0000-0002-1980-5293}}
\affiliation{Canadian Institute for Theoretical Astrophysics, University of Toronto, 60 St George St, Toronto, ON M5S 3H8, Canada}
\affiliation{David A. Dunlap Department of Astronomy and Astrophysics, University of Toronto, 50 St George St, Toronto ON M5S 3H4, Canada}
\affiliation{Department of Physics, 60 St George St, University of Toronto, Toronto, ON M5S 3H8, Canada}

\author{Carl L. Rodriguez\,\orcidlink{0000-0003-4175-8881}}
\affiliation{Department of Physics and Astronomy, University of North Carolina at Chapel Hill, 120 E. Cameron Ave, Chapel Hill, NC, 27599, USA}

\author{Will M. Farr\,\orcidlink{0000-0003-1540-8562}}
\affiliation{Department of Physics and Astronomy, Stony Brook University, Stony Brook, NY 11794, USA}
\affiliation{Center for Computational Astrophysics, Flatiron Institute, 162 Fifth Avenue, New York, NY 10010, USA}

\date{\today}

\begin{abstract}
\noindent
Mass distributions of binary black hole mergers inferred from gravitational-wave observations offer clues to how binaries form. These distributions conventionally label the heavier component as the primary, with mass \(m_1\), and the lighter as the secondary, with mass \(m_2\), such that \(m_1\geq m_2\). This ordering folds the physically labeled mass distribution along the equal-mass line, potentially reshaping the underlying population structure. In this Letter, we determine which population features can arise from mass ordering and which require genuine astrophysical structure. We show that ordering alone can produce (1) an \(m_2\) distribution that declines more rapidly than the \(m_1\) distribution, without intrinsic differences between the physically labeled components, and (2) an apparent concentration along equal masses, without correlations that preferentially pair black holes of equal mass. Moreover, different physically labeled populations can produce the same ordered mass distribution. Crucially, however, a preferred companion mass that remains constant as \(m_1\) increases requires a characteristic mass scale already present in the physically labeled population. Applied to binary black hole mergers in the fifth Gravitational-Wave Transient Catalog, our model recovers such a scale near \(m_2\simeq50\,\Msun\). Around this scale, the merger-rate density declines much more rapidly toward higher \(m_2\) than toward lower \(m_2\), ruling out an equal or weaker high-\(m_2\) decline at \(99.45\%\) credibility. This structure therefore requires a characteristic mass scale in at least one physically labeled component, but masses alone cannot determine which component carries it. Any formation model reproducing the observed high-mass binary black hole merger population must contain such a scale.
\end{abstract}

\maketitle

\noindent
Gravitational-wave (GW) observations of compact binary mergers provide a unique probe of the final stages of massive-star evolution. Predictions for the merger population depend sensitively on assumptions about poorly understood binary-evolution processes \citep{2020FrASS...7...38M,MandelFarmer2022,2022LRR....25....1M}. We therefore often begin by identifying features in the observed population, such as peaks, dips, and cutoffs, and then ask which astrophysical processes can robustly reproduce them.

One such set of recently identified features appears in the high-mass binary-black-hole merger population. Above \(m_1\gtrsim40\,\Msun\), the marginal distribution of the secondary mass, \(m_2\), declines more steeply than that of the primary mass, \(m_1\) \citep{2026arXiv260527226T}. The ridge of the joint merger-rate density in the \(m_1\)--\(m_2\) plane follows the equal-mass boundary up to \(m_1\simeq m_2\simeq45\,\Msun\), before turning and becoming approximately parallel to the \(m_1\) axis \citep{Tong2026,mould2026measurementprospectspairinstabilitymass}. Correspondingly, the mass-ratio distribution, with \(q=m_2/m_1\), is significantly flatter in this high-mass regime than in the rest of the population \citep{banagiri2025evidencesubpopulationsmergingbinary,2026arXiv260527226T}. Together with a change in the spin properties over the same mass range \citep{Antonini2025,antonini2026gravitationalwavesrevealpairinstability}, these features have been interpreted as evidence that first-generation--first-generation (1G--1G) mergers are suppressed at high masses and that second-generation--first-generation (2G--1G) mergers begin to dominate \citep{2022PhRvD.106j3013M,2024arXiv240601679P,Tong2026,antonini2026gravitationalwavesrevealpairinstability,afroz2025binaryblackholephase}. Under this interpretation, \(\sim45\,\Msun\) represents the maximum mass of a first-generation black hole produced through stellar evolution, marking the lower edge of the pair-instability mass gap \citep{FowlerHoyle1964,BarkatRakavy1967,RakavyShaviv1967,Fraley1968,Woosley_2002,Woosley_2021}.

This interpretation, however, may be affected by the convention used to label the binary components. The gravitational waveform is invariant under the simultaneous exchange of all parameters associated with the two components and therefore cannot identify which observed mass corresponds to which physically labeled black hole \citep{2019PhRvD.100h4041B,2021PhRvL.126q1103B,Sadiq_2023,2025PhRvL.134l1402G}. To separate physical identities from mass ordering, we introduce labeled masses \(m_{\rm A}\) and \(m_{\rm B}\) before defining the heavier and lighter components. The labels \(A\) and \(B\) represent any physical criterion used to distinguish the components. In isolated binary evolution, for example, they may identify the black holes descended from the initially heavier and lighter zero-age-main-sequence (ZAMS) stars. Binary interactions can reverse their mass ordering, so \(m_{\rm A}\) may be greater than, less than, or equal to \(m_{\rm B}\) at merger \citep{2022ApJ...933...86Z,Broekgaarden_2022,2022MNRAS.517.2738M,Olejak_2024,2026arXiv260523083G,2026arXiv260521580S,2026arXiv260626262C,2026arXiv260628506M,2026arXiv260627852H}.

GW observations recover only the ordered masses
\begin{equation*}
m_1=\max(m_{\rm A},m_{\rm B}), \qquad
m_2=\min(m_{\rm A},m_{\rm B}),
\end{equation*}
which enforce \(m_1\geq m_2\). This map folds the labeled \(m_{\rm A}\)--\(m_{\rm B}\) plane along \(m_{\rm A}=m_{\rm B}\) and maps both halves onto the ordered \(m_1\)--\(m_2\) plane. This folding changes the geometry of the two-dimensional distribution, so a structure in the ordered plane need not have the same form in the underlying labeled-mass plane. It can also make distinct labeled subpopulations overlap in the ordered plane. Once they overlap, the \(m_1\)--\(m_2\) distribution alone cannot separate them unless they retain some distinguishing property.

\begin{figure}[h!]
    \centering
    \includegraphics[width=0.86\textwidth]{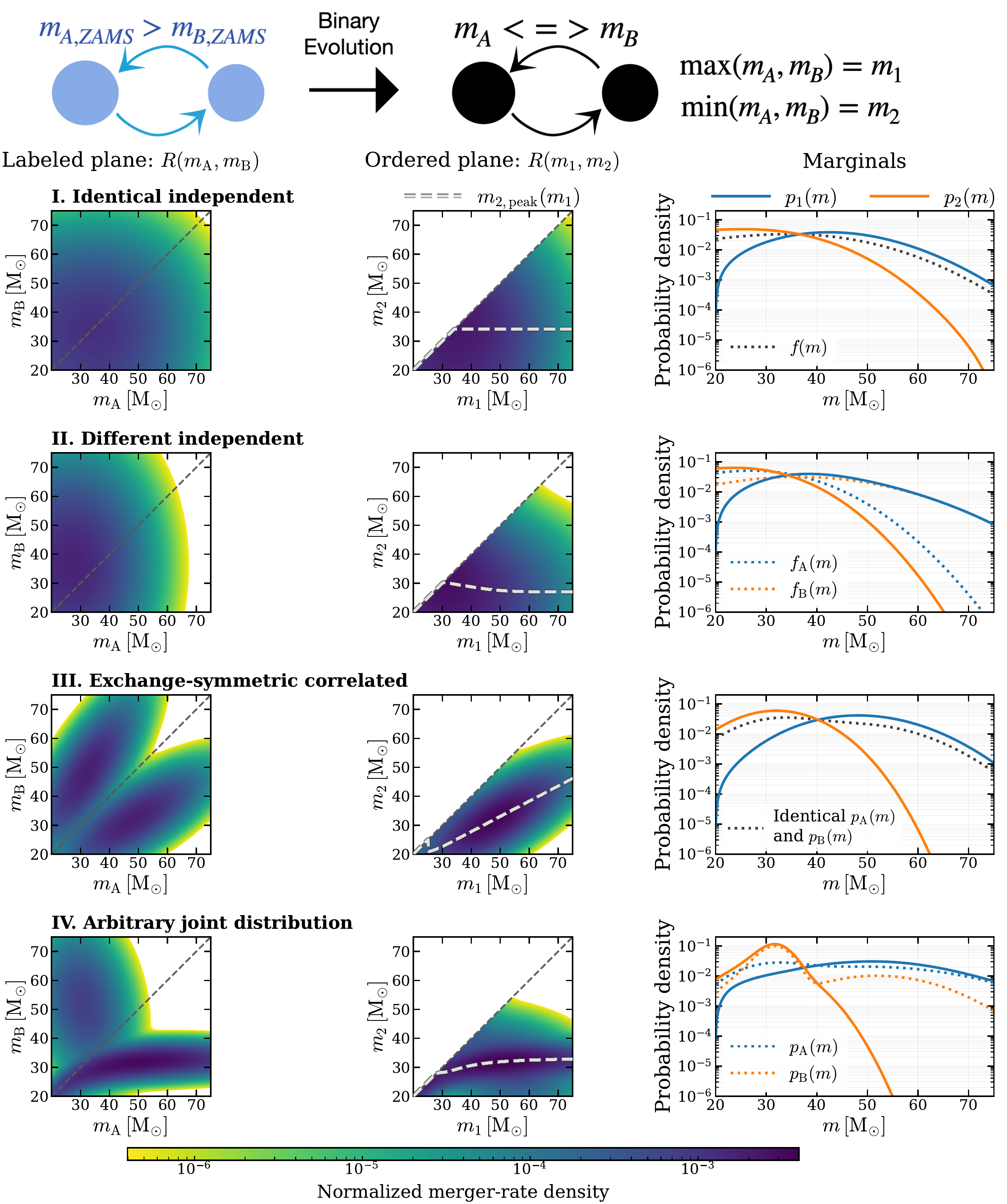}
    \caption{
        Mass ordering can create apparent features in the inferred astrophysical mass distribution. The physically labeled merger masses \((m_{\rm A},m_{\rm B})\) map to the ordered masses \(m_1=\max(m_{\rm A},m_{\rm B})\) and \(m_2=\min(m_{\rm A},m_{\rm B})\). Rows I--IV show identical independent, different independent, exchange-symmetric correlated, and arbitrary joint distributions in \((m_{\rm A},m_{\rm B})\). For Rows II--IV, where a steeper decline of the ordered \(m_2\) marginal is not guaranteed, we deliberately choose examples in which the \(m_2\) marginal declines more steeply than the \(m_1\) marginal, even though the labeled marginals either decline identically or the \(m_{\rm A}\) marginal declines more steeply than the \(m_{\rm B}\) marginal. Ordering can generate apparent correlations between \(m_1\) and \(m_2\), produce different \(p_1(m)\) and \(p_2(m)\) marginals, and substantially reshape features in the labeled distribution. The black dotted line marks the maximum of the conditional distribution \(p(m_2\mid m_1)\). The functional forms and code used to produce this figure are available \href{https://github.com/SoumendraRoy/Mass_Ordering/blob/main/scripts/Plots_Main.ipynb}{here}.
    }
    \label{fig:ordering-cartoon}
\end{figure}

\paragraph{\textbf{Mass Ordering and Pairing.}} To determine which features arise from this folding and which require genuine structure in the labeled-mass population, we consider four increasingly general pairing cases \citep{2009A&A...493..979K,Picky_Partners_Fishbach_2020,farah2024kindcomparingbigsmall,2026arXiv260210186G}, illustrated in Fig.~\ref{fig:ordering-cartoon} of the End Matter:

\paragraph{Case I: Identical independent labeled masses.}
We begin with the simplest case, in which \(m_{\rm A}\) and \(m_{\rm B}\) are drawn independently from the same normalized distribution \(f\) \citep{2001MNRAS.321..149M}. The ordered joint distribution is
\begin{equation}
P(m_1,m_2)
=
2f(m_1)f(m_2)\Theta(m_1-m_2),
\label{eq:ordering_case_I_joint}
\end{equation}
where \(\Theta\) is the Heaviside step function. The factor of two accounts for the two labeled configurations that map to the same ordered pair. The two-dimensional structure is therefore determined entirely by the common one-dimensional distribution \(f\) and the ordering boundary \(m_1\geq m_2\).

Let \(F(m)\) denote the cumulative distribution function of \(f(m)\). The ordered marginals are
\begin{equation}
p_1(m)=2f(m)F(m),\qquad
p_2(m)=2f(m)\bigl[1-F(m)\bigr].
\label{eq:ordering_case_I_marginals}
\end{equation}
Their logarithmic slopes,
\(s_i(m)={\rm d}\log p_i(m)/{\rm d}\log m\), satisfy
\begin{equation}
s_2(m)-s_1(m)
=
-\frac{m f(m)}
{F(m)\bigl[1-F(m)\bigr]}
<0.
\label{eq:ordering_case_I_slope_difference}
\end{equation}
Thus, wherever both ordered marginals decline, the \(m_2\) marginal declines more steeply than the \(m_1\) marginal, irrespective of the form of \(f\). A steeper high-mass \(m_2\) decline can therefore arise purely from mass ordering, without any intrinsic difference between the physically labeled black holes.

Now suppose that \(f\) peaks at \(m_{\rm peak}\). When \(m_1<m_{\rm peak}\), the preferred value \(m_2=m_{\rm peak}\) lies outside the allowed region \(m_2\leq m_1\), so the conditional density peaks on the equal-mass boundary. Once \(m_1>m_{\rm peak}\), this preferred value becomes accessible and the ridge follows \(m_2=m_{\rm peak}\). The conditional mass-ratio distribution therefore peaks at
\begin{equation}
q_{\rm peak}(m_1)
=
\begin{cases}
1, & m_1\leq m_{\rm peak},\\[2pt]
m_{\rm peak}/m_1, & m_1>m_{\rm peak}.
\end{cases}
\label{eq:ordering_case_I_q_peak}
\end{equation}
The equal-mass segment is produced by the ordering boundary. By contrast, the constant-\(m_2\) branch inherits the characteristic scale \(m_{\rm peak}\) already present in the labeled-mass distribution. In terms of mass ratio, this pre-existing constant mass appears as \(q_{\rm peak}\propto m_1^{-1}\). 

\paragraph{Case II: Different independent labeled masses.}
We next draw \(m_{\rm A}\) and \(m_{\rm B}\) independently from different normalized distributions, \(f_{\rm A}\) and \(f_{\rm B}\). The ordered distribution is
\begin{equation}\label{eq:ordering_case_II_joint}
\begin{aligned}
P(m_1,m_2)
&=
\bigl[
f_{\rm A}(m_1)f_{\rm B}(m_2)
+
f_{\rm B}(m_1)f_{\rm A}(m_2)
\bigr] \\
&\quad\times\Theta(m_1-m_2).
\end{aligned}
\end{equation}
The two terms correspond to the two possible assignments of the physical labels to the same ordered binary. Independence restricts intrinsic structures in the labeled-mass plane to branches parallel to either mass axis, and folding superposes these branches with their reflected counterparts. Any constant-component-mass branch therefore reflects a characteristic scale already present in \(f_{\rm A}\) or \(f_{\rm B}\), rather than one created by ordering. If such a branch lies outside the allowed region \(m_2\leq m_1\), the conditional ridge instead follows the equal-mass boundary, producing an apparent equal-mass ridge without intrinsic correlation.

Unlike Case I, ordering does not determine which ordered marginal declines more steeply. Defining
\begin{equation*}
F_{\rm A}(m)=\int_{m_{\rm low}}^{m}f_{\rm A}(t)\,{\rm d}t,
\qquad
F_{\rm B}(m)=\int_{m_{\rm low}}^{m}f_{\rm B}(t)\,{\rm d}t,
\end{equation*}
the \(m_2\) marginal declines more steeply than the \(m_1\) marginal only if
\begin{equation}\label{eq:ordering_case_II_slope_condition}
\bigl(F_{\rm A}-F_{\rm B}\bigr)
\left(
\frac{{\rm d}\log f_{\rm A}}{{\rm d}m}
-
\frac{{\rm d}\log f_{\rm B}}{{\rm d}m}
\right)
<
2\bigl(f_{\rm A}+f_{\rm B}\bigr).
\end{equation}
The relative slopes therefore depend on both labeled distributions rather than on mass ordering alone. In particular, a more rapidly declining \(m_2\) marginal does not by itself imply a cutoff or a characteristic \(m_2\) scale in the joint \(m_1\)--\(m_2\) distribution.

\textit{Case III: Exchange-symmetric labeled masses.} 
When no physical criterion distinguishes the two components, assigning the labels \(A\) and \(B\) randomly with equal probability produces an exchange-symmetric labeled distribution,
\begin{equation}
    f(m_{\rm A},m_{\rm B})=f(m_{\rm B},m_{\rm A}).
\end{equation}
For example, in dynamically assembled binary black hole mergers in globular clusters, both black holes are drawn from the same population containing black holes of multiple generations and assembled through an exchange-symmetric pairing function, which can introduce correlations between their masses \citep{Rodriguez_2016, 2017MNRAS.469.4665P, Rodriguez2019, 2026ApJ...997..267Y, 2024ApJ...967...62Y}.

The corresponding ordered distribution is
\begin{equation}\label{eq:ordering_case_III_joint}
P(m_1,m_2)
=
2f(m_1,m_2)\Theta(m_1-m_2).
\end{equation}
In the independent cases, preferred component masses produce branches parallel to the mass axes. When the labeled masses are correlated, however, the preferred value of one mass can vary with the other, producing a tilted or curved ridge. After folding, the corresponding conditional mass-ratio peak is therefore not restricted to \(q_{\rm peak}=1\) or \(q_{\rm peak}\propto m_1^{-1}\). For example, a scale-free correlation of the form \(m_2=q_0m_1\) produces \(q_{\rm peak}=q_0<1\). This introduces a preferred mass ratio but no absolute mass scale, as shown in Supplemental Material \hyperref[app:pq]{I}.

Correlations also make the ordered marginal slopes depend on the full two-dimensional distribution. Even though exchange symmetry makes the labeled marginals identical, either ordered marginal can decline more steeply. This reinforces the conclusion of Case II that the relative slopes of the ordered marginals alone cannot establish the presence of a cutoff or characteristic \(m_2\) scale in the joint \(m_1\)--\(m_2\) distribution.

\paragraph{Case IV: Arbitrary joint labeled masses.}
Finally, we allow the labeled masses to follow an arbitrary joint distribution \(f(m_{\rm A},m_{\rm B})\), without assuming independence or exchange symmetry. The ordered distribution is
\begin{equation}
    \label{eq:ordering_case_IV_joint}
    P(m_1,m_2)
    =
    \bigl[
        f(m_1,m_2)+f(m_2,m_1)
    \bigr]\Theta(m_1-m_2).
\end{equation}
The ordering map only exchanges and folds the two mass coordinates and introduces no dimensional parameter. If the labeled distribution is scale free under a simultaneous rescaling of both masses, Eq.~\eqref{eq:ordering_case_IV_joint} preserves the same scaling. Ordering can therefore create structure along the scale-free boundary \(m_1=m_2\), but it cannot generate an interior characteristic mass scale absent from the labeled distribution. Subject to this restriction, the merger-rate geometry, the relative slopes of the ordered marginals, and the peak of \(p(q\mid m_1)\) can take arbitrary forms and depend on the full labeled joint distribution. Features of the ordered marginals or a preference for equal masses therefore establish neither intrinsic differences between \(m_{\rm A}\) and \(m_{\rm B}\) nor correlations between them. Intrinsic label differences remain non-identifiable from ordered masses, while correlations require structure in the joint distribution that cannot be produced by ordering the labeled masses.

Reconstructing the labeled-mass distribution from the ordered distribution requires assumptions about the labeled population. Under the assumptions of Cases I--III, the labeled distribution can be recovered uniquely or up to exchanging \(A\) and \(B\); we perform these reconstructions explicitly in Supplemental Material \hyperref[app:labeled_reconstruction]{II}. Without these restrictions, the ordered distribution determines only the exchange-symmetric combination \(f(m_{\rm A},m_{\rm B})+f(m_{\rm B},m_{\rm A})\). Any permitted exchange-antisymmetric component leaves \(P(m_1,m_2)\) unchanged, so infinitely many labeled populations are consistent with the same ordered distribution. Masses alone therefore cannot recover the full merger-rate density in the labeled-mass plane or determine whether an observed feature is associated with \(A\) or \(B\).

\paragraph{\textbf{Consequences of Mass Ordering.}} The four cases, illustrated in Fig.~\ref{fig:ordering-cartoon} of the End Matter, lead to three main conclusions:
\begin{enumerate}
    \item \textbf{Marginal slopes can be artifacts of ordering.} Mass ordering can make the \(m_2\) marginal decline more steeply than the \(m_1\) marginal. This difference alone neither demonstrates an intrinsic distinction between the labeled components nor establishes a cutoff or characteristic \(m_2\) scale in the joint distribution.
    
    \item \textbf{Equal-mass ridges can be artifacts of ordering.} Mass ordering can produce an apparent ridge along the equal-mass line \(m_1=m_2\), even when the physically labeled masses are independent.
    
    \item \textbf{Axis-parallel ridges require astrophysical mass scales.} An axis-parallel ridge provides evidence for a characteristic component-mass scale already present in the labeled population. Mass ordering can obscure whether \(A\) or \(B\) carries this scale, but it cannot generate the scale from an underlying scale-free distribution.
\end{enumerate}
In Supplemental Material \hyperref[app:posterior_vs_prior]{IV}, we clarify why prior-to-posterior predictive changes cannot, by themselves, identify astrophysical features in the underlying labeled population.

\begin{figure*}[t]
    \centering
    \includegraphics[width=\textwidth]{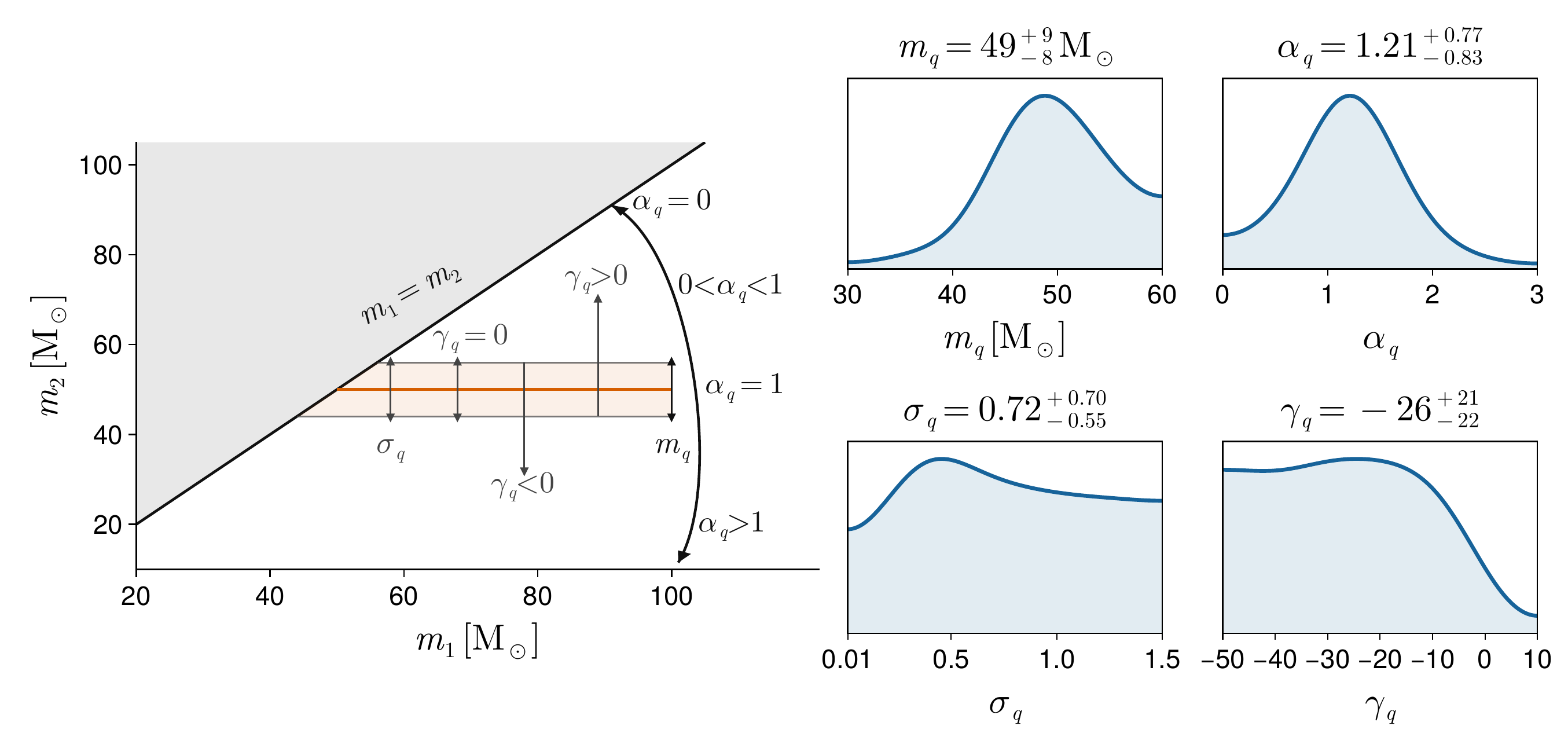}
    \caption{
        Geometry and inferred parameters of the conditional mass-ratio model for the high-mass population. The left panel illustrates how \(m_q\), \(\alpha_q\), \(\sigma_q\), and \(\gamma_q\) control the characteristic mass scale, ridge geometry, width, and skewness, respectively. The right panels show their marginalized posterior distributions. The inferred \(\alpha_q=1.21^{+0.77}_{-0.83}\) is consistent with unity, indicating that the characteristic companion-mass scale in \(m_2\) stays approximately constant as \(m_1\) increases, with \(m_q=49^{+9}_{-8}\,\Msun\). Under the uniform priors \(\sigma_q\sim\mathrm{U}(0.01,1.5)\) and \(\gamma_q\sim\mathrm{U}(-50,10)\), the inferred width \(\sigma_q=0.72^{+0.70}_{-0.55}\) is bounded away from its lower prior boundary, while posterior support for \(\gamma_q \geq 0\) is strongly suppressed. Together, these constraints imply a sharper decline toward higher \(m_2\) than toward lower \(m_2\), showing a strong suppression of the high-\(m_2\) tail. The full joint posterior is shown in Fig.~\ref{fig:high_mass_corner} of Supplemental Material \hyperref[app:population_inference]{III}. Code used to produce this figure is available \href{https://github.com/SoumendraRoy/Mass_Ordering/blob/main/scripts/Plots_Main.ipynb}{here}.
        }
    \label{fig:q_model_summary}
\end{figure*}

\begin{figure*}[t]
    \centering
    \includegraphics[
        width=\textwidth,
        keepaspectratio
    ]{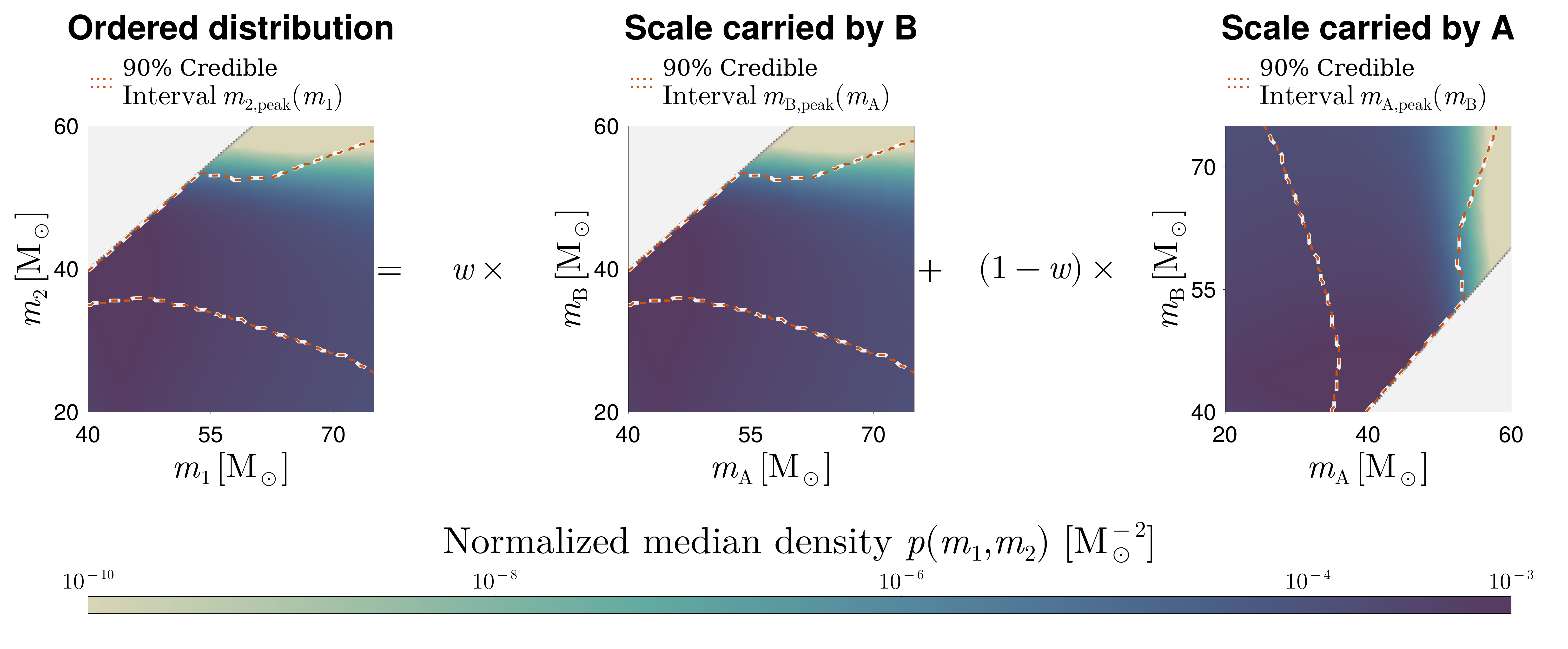}
    \caption{
        The GWTC-5 high-mass BBH merger population requires a characteristic mass scale in the physically labeled population, but does not determine which component carries it. The left panel shows the posterior median ordered mass distribution, with each draw normalized over \(5\,\Msun<m_2<m_1<300\,\Msun\). Dotted curves bound the 90\% credible interval of \(m_{2,\rm peak}(m_1)\), the secondary mass maximizing the merger-rate density at fixed \(m_1\), with reflected counterparts in the labeled panels. The middle and right panels assign the distribution to the \(m_{\rm A}\geq m_{\rm B}\) and \(m_{\rm B}\geq m_{\rm A}\) branches with weights \(w\) and \(1-w\), where \(0\leq w\leq1\). Constant weights preserve the horizontal and vertical ridges shown; \(w=1\) assigns the scale to \(B\), and \(w=0\) to \(A\). Mass-dependent weights can reshape the labeled ridges without changing the ordered distribution or its required mass scale. Pointwise density uncertainties and plotting code are available \href{https://github.com/SoumendraRoy/Mass_Ordering/blob/main/scripts/Plots_Main.ipynb}{here}.
    }
    \label{fig:ordering-degeneracy}
\end{figure*}

\paragraph{\textbf{The High-Mass Binary Black Hole Merger Population.}} We now interpret the high-mass structure of the binary black hole (BBH) merger population in light of these conclusions. Using BBH events from the fifth Gravitational-Wave Transient Catalog (GWTC-5) \citep{2026arXiv260527223T,2026arXiv260527224T,2026arXiv260527225T}, observed by the LIGO \citep{aligo}, Virgo \citep{avirgo}, and KAGRA \citep{10.1093/ptep/ptaa125} detectors, we infer the merger-rate density in the ordered \((m_1,m_2)\) plane through hierarchical Bayesian inference that accounts for measurement uncertainties and selection effects. Writing \(q=m_2/m_1\), we parametrize the merger-rate density in the high-mass regime as
\begin{equation}\label{eq:mass_model}
    \frac{d^2\mathcal{R}}{dm_1\,dq}
    =
    R_0\,p(m_1)\,p(q\mid m_1),
\end{equation}
where \(p(m_1)\propto m_1^{-\alpha}\) and
\begin{equation}
\begin{aligned}
    p(q\mid m_1)
    &=
    \operatorname{SkewNormal}_{[5\Msun/m_1,\,1]}
    \Biggl[
        q\,\Bigm| \\[-0.2em]
    &\qquad
        \xi_q=
        \left(\frac{m_q}{m_1}\right)^{\alpha_q},
        \ \sigma_q,\ \gamma_q
    \Biggr].
\end{aligned}
\end{equation}
Here, \(\sigma_q\) and \(\gamma_q\) control the width and skewness, while \(\alpha_q\) determines the geometry of the merger-rate ridge through
\begin{equation}\label{eq:m2loc}
    m_{2,\mathrm{loc}}
    =
    m_1\xi_q
    =
    m_q^{\alpha_q}m_1^{1-\alpha_q}.
\end{equation}
Thus, \(\alpha_q=0\) follows the equal-mass boundary, whereas \(\alpha_q=1\) produces a ridge parallel to the \(m_1\) axis at the characteristic \(m_2\) scale \(m_q\). For \(0<\alpha_q<1\), the ridge lies between the equal-mass and constant-\(m_2\) tracks, while for \(\alpha_q>1\), it decreases with \(m_1\). Fig.~\ref{fig:q_model_summary} illustrates how these parameters control the ridge geometry. We describe the complete population model and hierarchical likelihood in Supplemental Material \hyperref[app:population_inference]{III}.

As shown in Fig.~\ref{fig:q_model_summary}, we infer \(\alpha_q=1.2^{+0.8}_{-0.8}\) and \(m_q=49^{+9}_{-8}\Msun\), where the uncertainties denote 90\% credible intervals. The inferred \(\alpha_q\) is consistent with unity, implying that the preferred secondary-mass scale is nearly independent of \(m_1\). Correspondingly, the ordered merger-rate density in Fig.~\ref{fig:ordering-degeneracy} exhibits a high-mass ridge whose conditional maximum, \(m_{2,\mathrm{peak}}(m_1)\), becomes approximately parallel to the \(m_1\) axis. The GWTC-5 population therefore supports a characteristic component-mass scale near \(m_2\simeq50\Msun\). Supplemental Material \hyperref[app:population_inference]{III} presents the full joint posterior, including correlations among the hyperparameters, in Fig.~\ref{fig:high_mass_corner}, and compares the inferred mass distribution with the GWTC-5 \texttt{PixelPop} analysis in Fig.~\ref{fig:gwtc5_comparison}.

The existence of a characteristic scale near \(m_2\simeq50\,\Msun\), however, does not establish a gap in \(m_2\). \cite{Tong2026} reported evidence for a sharp secondary-mass gap, whereas \cite{Ray_2026} found that the data do not require an empty gap and instead favor a smooth falloff when analyzed with a more flexible mass model. To revisit this question with GWTC-5, we compare the decline in the merger-rate density on either side of the \(m_2\simeq50\,\Msun\) ridge. As shown in Fig.~\ref{fig:q_model_summary}, the inferred finite width (\(\sigma_q\)) and strong preference for negative skewness (\(\gamma_q\)) produce a much sharper decline toward higher \(m_2\) than toward lower \(m_2\). At \(m_1=70\,\Msun\), the decline between \(m_2=50\,\Msun\) and \(70\,\Msun\) is at least \(9.63\) times stronger than that between \(m_2=50\,\Msun\) and \(30\,\Msun\) at \(90\%\) credibility, while an equal or weaker high-\(m_2\) decline is disfavored at \(99.45\%\) credibility. However, under the prior \(\gamma_q\sim U(-50,10)\), the posterior becomes nearly constant for \(\gamma_q\lesssim-10\), indicating that the data favor a sharp high-\(m_2\) falloff but cannot distinguish a steep decline from an abrupt cutoff. The GWTC-5 data therefore support an asymmetric suppression toward high \(m_2\), while remaining insensitive to whether it forms a strictly empty gap or a sufficiently steep continuous decline. We check the consistency of this result with the event-level mass posteriors in Figs.~\ref{fig:event_reweighting_1} and \ref{fig:event_reweighting_2} of Supplemental Material \hyperref[app:population_inference]{III}. We test the robustness of the inference in Figs.~\ref{fig:pcut_robustness} and \ref{fig:lower_mass_sensitivity} of Supplemental Material \hyperref[app:population_inference]{III}.

Regardless of whether the high-\(m_2\) suppression forms a hard gap or a steep continuous falloff, the axis-parallel ridge requires a characteristic mass scale in the physically labeled population. Fig.~\ref{fig:ordering-degeneracy} illustrates the remaining ambiguity: \(w\) is the fraction of the density at each ordered mass pair assigned to the branch \(m_{\rm A}\geq m_{\rm B}\), with the remainder assigned to \(m_{\rm B}\geq m_{\rm A}\). In the \(m_{\rm A}\)--\(m_{\rm B}\) plane, \(w=1\) means that the preferred companion always carries label \(B\), producing a horizontal ridge at nearly constant \(m_{\rm B}\); \(w=0\) means that it always carries label \(A\), producing a vertical ridge at nearly constant \(m_{\rm A}\). These labeled ridges can be reshaped by mass-dependent weights without changing the ordered distribution, as explained in Supplemental Material~\hyperref[app:ridge_labels]{V}. These different assignments can change the labeled ridge geometry, but cannot remove the characteristic mass scale from the underlying population: over a region of the \(m_A\)--\(m_B\) plane, either \(m_A\) or \(m_B\) must lie preferentially near \(50\,\Msun\), so that \(m_2=\min(m_A,m_B)\) remains near \(50\,\Msun\) for a range of \(m_1\). The characteristic mass scale required by the GWTC-5 high-mass BBH population is therefore robust to mass ordering, but masses alone cannot determine whether \(A\) or \(B\) carries it.

Breaking this degeneracy requires observables with label-dependent population structure. If \(A\) and \(B\) have distinct spin distributions or belong to subpopulations with different redshift evolution, a joint mass--spin--redshift analysis could constrain \(w\) and associate the characteristic scale with one of the labeled components \citep[see, e.g.,][Hussain et al.~(in preparation)]{2021PhRvL.126q1103B,2025PhRvL.134l1402G,2026CQGra..43h5005S}. Without such additional information, exchanging \(A\) and \(B\) leaves the ordered mass distribution unchanged.

Alternatively, we can avoid assigning physical meaning to the ordered labels by analyzing exchange-symmetric variables, such as the chirp mass and symmetric mass ratio, and forward-modeling formation-channel simulations into these observables \citep{2024MNRAS.527..298T,2025ApJ...995..177T,2025arXiv251025579T}. However, uncertainties in binary-evolution physics allow the predictions to vary widely. Identifying the structure required in physically labeled population quantities can therefore guide which physical ingredients successful models must include. Our result provides one such requirement: any model reproducing the high-mass BBH merger population must contain a characteristic mass scale near \(50\,\Msun\) in at least one of its components.

\paragraph{\textbf{Declaration.}} After completing this work, we became aware of a similar analysis carried out independently at the same time in \cite{2026arXiv261009834A}. The ordered-draw construction studied there, in which the two component masses are independently drawn from a common distribution and then ordered, corresponds to Case I of our framework.

\paragraph{\textbf{Data Release.}} The analysis code required to reproduce all figures and numerical results in this paper is publicly available on \href{https://github.com/SoumendraRoy/Mass_Ordering}{GitHub} under the GNU General Public License. We use parameter-estimation samples for the GWTC-5 events \citep{ligo_scientific_collaboration_and_virgo_2026_20276106, ligo_scientific_collaboration_and_virgo_2026_20291740, ligo_scientific_collaboration_and_virgo_2026_20275769, ligo_scientific_collaboration_and_virgo_2023_8177023, ligo_scientific_collaboration_and_virgo_2022_6513631} and the corresponding sensitivity estimates \citep{ligo_scientific_collaboration_2026_19500052}. We also use the GWTC-5 population data products \citep{ligo_scientific_collaboration_2026_20292639} to compare our results with the GWTC-5 population analysis \citep{2026arXiv260527226T} in Fig.~\ref{fig:gwtc5_comparison} of Supplemental Material \hyperref[app:population_inference]{III}.

\paragraph{\textbf{Software.}} \texttt{Turing.jl} \citep{ge2018t}; \texttt{Makie.jl} \citep{DanischKrumbiegel2021}; \texttt{PopModels.jl} \citep{popmodel}.

\paragraph{\textbf{Acknowledgments.}} We thank Michael Zevin for the LIGO P\&P review. We thank Lucas de S\'a and Reed Essick for helpful discussions, Claire Ye for pointing us to relevant references, Anarya Ray for comments on high-\(m_2\) falloff, and Jack Heinzel for running the \texttt{PixelPop} analysis and allowing us to use its results in Supplemental Material \hyperref[app:population_inference]{III}. We are grateful to members of CCA, CITA, and the LVK R\&P group for many stimulating discussions that improved this work.

SKR thanks the Center for Computational Astrophysics at the Flatiron Institute for hospitality while this research was carried out. The Center for Computational Astrophysics at the Flatiron Institute is supported by the Simons Foundation. The computations in this work were, in part, run at facilities supported by the Scientific Computing Core at the Flatiron Institute, a division of the Simons Foundation. This material is based upon work supported by NSF’s LIGO Laboratory which is a major facility fully funded by the National Science Foundation. LvS is supported by VI.Veni.242.115, Grant ID \url{https://doi.org/10.61686/XVIAV86753}. MF acknowledges support from the Natural Sciences and Engineering Research Council of Canada under grant RGPIN-2023-05511 and the Ontario Early Researcher Award ER24-18-170.

This paper carries the LIGO report number P2600544.

\bibliographystyle{apsrev4-2-short}
\bibliography{main}


\clearpage
\onecolumngrid

\section*{End Matter}

\clearpage
\twocolumngrid


\section*{Supplemental Material I: Reading mass scales from the conditional mass-ratio distribution}
\label{app:pq}

At fixed $m_1$, changing variables from $m_2$ to $q=m_2/m_1$ gives
\begin{equation}
    p(q\mid m_1)
    =
    \frac{
        m_1 P(m_1,qm_1)
    }{
        \displaystyle
        \int_{m_{\min}}^{m_1}
        P(m_1,m_2)\,dm_2
    },
    \qquad
    \frac{m_{\min}}{m_1}\leq q\leq1.
    \label{eq:pq_general}
\end{equation}

\paragraph{Case I.}
For identical independent labeled masses,
\begin{equation}
    p_{\rm I}(q\mid m_1)
    \propto
    f(qm_1).
    \label{eq:pq_case1}
\end{equation}
If $f(m)\propto m^\gamma$, then
\begin{equation}
    p_{\rm I}(q\mid m_1)\propto q^\gamma.
\end{equation}
An increasing power law therefore peaks at $q=1$, while a decreasing power law peaks at the lower allowed value of $q$. A flat mass distribution gives a flat conditional mass-ratio distribution.

For a Gaussian \(f\) with mean \(\mu\) and width \(\sigma\), the conditional mass-ratio distribution is
\begin{equation}
p_{\rm I}(q\mid m_1)
=
\mathcal{N}_{[m_{\min}/m_1,\,1]}
\left(q\,\middle|\,\frac{\mu}{m_1},\frac{\sigma}{m_1}\right),
\end{equation}
where \(\mathcal{N}_{[a,b]}\) denotes a truncated Gaussian between \(a\) and \(b\). Its peak lies at
\begin{equation}
q_{\rm peak}(m_1)
=
\begin{cases}
1, & m_1\leq\mu,\\[3pt]
\mu/m_1, & m_1>\mu.
\end{cases}
\label{eq:qpeak_case1}
\end{equation}
The peak remains on the equal-mass boundary until the preferred mass $\mu$ becomes accessible. It then follows a constant-$m_2$ branch.

\paragraph{Case II.}
For different independent labeled masses,
\begin{equation}
    p_{\rm II}(q\mid m_1)
    \propto
    f_A(m_1)f_B(qm_1)
    +
    f_B(m_1)f_A(qm_1).
    \label{eq:pq_case2}
\end{equation}
For power laws $f_A(m)\propto m^\alpha$ and
$f_B(m)\propto m^\beta$,
\begin{equation}
    p_{\rm II}(q\mid m_1)
    \propto
    q^\alpha+q^\beta.
    \label{eq:pq_case2_powerlaw}
\end{equation}
If both distributions increase, the mode is at $q=1$. If both decrease, it is at the lower boundary. When one increases and the other decreases, the interior stationary point,
\begin{equation}
    q_{\rm turn}
    =
    \left(
        \frac{-\beta}{\alpha}
    \right)^{1/(\alpha-\beta)},
\end{equation}
is a minimum, so the maxima again lie at the boundaries.

If $f_A$ and $f_B$ instead have preferred masses $\mu_A$ and $\mu_B$, they can produce interior branches near
\begin{equation}
    q\simeq\frac{\mu_A}{m_1},
    \qquad
    q\simeq\frac{\mu_B}{m_1},
    \label{eq:qpeak_case2}
\end{equation}
when those masses lie within the allowed range. Depending on their relative contributions, the conditional distribution may contain one or two such peaks.

\paragraph{Case III.}
A correlated distribution can produce behavior that is impossible for independent scale-free masses. Consider an exchange-symmetric, scale-free joint distribution satisfying
\begin{equation}
    f(m_A,m_B)=f(m_B,m_A)
\end{equation}
and
\begin{equation}
    f(\lambda m_A,\lambda m_B)
    =
    \lambda^\kappa f(m_A,m_B).
\end{equation}
Using $f(m_1,qm_1)=m_1^\kappa f(1,q)$, its conditional mass-ratio
distribution is
\begin{equation}
    p_{\rm III}(q\mid m_1)
    =
    \frac{
        f(1,q)
    }{
        \displaystyle
        \int_{m_{\min}/m_1}^{1}f(1,u)\,du
    }.
    \label{eq:pq_case3_scalefree}
\end{equation}
Its shape is therefore independent of $m_1$ except for the changing lower boundary, and a correlated ridge can remain at a constant mass ratio,
\begin{equation}
    q_{\rm peak}(m_1)=q_0<1.
\end{equation}
Such a feature introduces a preferred mass ratio, but not a characteristic mass scale.

More generally, let the ridge follow
\begin{equation}
    m_{2,\rm peak}(m_1)=r(m_1).
\end{equation}
Its location in mass ratio is
\begin{equation}
    q_{\rm peak}(m_1)
    =
    \frac{r(m_1)}{m_1}.
    \label{eq:qpeak_ridge}
\end{equation}
A ridge with $r(m_1)=q_0m_1$ gives a constant $q_{\rm peak}$, whereas a ridge with $r(m_1)=\mu$ gives
\begin{equation}
    q_{\rm peak}(m_1)=\frac{\mu}{m_1}.
    \label{eq:qpeak_fixed_mass}
\end{equation}

\paragraph{Case IV.}
For an arbitrary labeled distribution, define its exchange-symmetric part as
\begin{equation}
    f_+(m_A,m_B)
    =
    \frac{
        f(m_A,m_B)+f(m_B,m_A)
    }{2}.
\end{equation}
The conditional mass-ratio distribution depends only on this part:
\begin{equation}
    p_{\rm IV}(q\mid m_1)
    \propto
    f_+(m_1,qm_1).
    \label{eq:pq_case4}
\end{equation}
The exchange-antisymmetric part is therefore invisible in $p(q\mid m_1)$. It can change which physical label carries a mass scale, but not the observed location or shape of that feature.

The distinction is therefore sharp. A peak that moves away from $q=1$ does not by itself require a characteristic mass: a scale-free correlation can produce a constant $q_{\rm peak}<1$. By contrast,
\begin{equation}
    q_{\rm peak}(m_1)\simeq\frac{m_\ast}{m_1}
\end{equation}
implies
\begin{equation}
    m_{2,\rm peak}(m_1)
    =
    m_1q_{\rm peak}(m_1)
    \simeq m_\ast.
\end{equation}
This behavior directly requires a characteristic component-mass scale. Mass ordering can obscure whether this scale belongs to $m_A$ or $m_B$, but it cannot create the scale. The robust high-mass feature is therefore the flattening of $m_{2,\rm peak}(m_1)$, not the $q$ distribution peaking away from unity.

\section*{Supplemental Material II: Reconstructing Labeled Mass Plane from Ordered Mass Distribution}
\label{app:labeled_reconstruction}

Here we explicitly invert the ordering map under the assumptions of Cases I--III.

\paragraph{Case I.}
For identical independent labeled masses, Eq.~\eqref{eq:ordering_case_I_marginals} gives
\begin{equation}
    p_1(m)+p_2(m)=2f(m).
\end{equation}
The common labeled-mass distribution is therefore
\begin{equation}
    f(m)=\frac{p_1(m)+p_2(m)}{2},
    \label{eq:case1_reconstruction}
\end{equation}
which uniquely determines the labeled joint distribution
\begin{equation}
    f_{\rm I}(m_{\rm A},m_{\rm B})
    =
    f(m_{\rm A})f(m_{\rm B}).
\end{equation}

\paragraph{Case II.}
For different independent labeled masses, the cumulative distributions \(F_{\rm A}\) and \(F_{\rm B}\), defined in the main text, give the ordered marginals
\begin{align}
    p_1(m)
    &=
    f_{\rm A}(m)F_{\rm B}(m)
    +
    f_{\rm B}(m)F_{\rm A}(m),
    \label{eq:case2_ordered_p1}
    \\
    p_2(m)
    &=
    f_{\rm A}(m)\bigl[1-F_{\rm B}(m)\bigr]
    +
    f_{\rm B}(m)\bigl[1-F_{\rm A}(m)\bigr].
    \label{eq:case2_ordered_p2}
\end{align}
Define their cumulative distributions as
\begin{equation}
    G_i(m)
    =
    \int_{m_{\min}}^{m}p_i(t)\,dt,
    \qquad
    S(m)=G_1(m)+G_2(m).
\end{equation}
Equations~\eqref{eq:case2_ordered_p1} and
\eqref{eq:case2_ordered_p2} then give
\begin{equation}
    G_1(m)=F_{\rm A}(m)F_{\rm B}(m),
    \qquad
    S(m)=F_{\rm A}(m)+F_{\rm B}(m).
\end{equation}
Thus, \(F_{\rm A}(m)\) and \(F_{\rm B}(m)\) are the two roots of
\begin{equation}
    x^2-S(m)x+G_1(m)=0,
\end{equation}
and hence
\begin{align}
    F_{\rm A}(m)
    &=
    \frac{
        S(m)\pm\sqrt{S^2(m)-4G_1(m)}
    }{2},
    \label{eq:case2_FA_reconstruction}
    \\
    F_{\rm B}(m)
    &=
    \frac{
        S(m)\mp\sqrt{S^2(m)-4G_1(m)}
    }{2}.
    \label{eq:case2_FB_reconstruction}
\end{align}
Differentiating these expressions gives \(f_{\rm A}\) and \(f_{\rm B}\), and therefore
\begin{equation}
    f_{\rm II}(m_{\rm A},m_{\rm B})
    =
    f_{\rm A}(m_{\rm A})f_{\rm B}(m_{\rm B}).
\end{equation}
The two signs in these expressions recover the cumulative distributions $F_A$ and $F_B$, but do not specify which belongs to which physical label. If the two distributions cross, their assignment to $A$ and $B$ could change across the crossing without altering the ordered marginals. To resolve this ambiguity, the reconstructed distributions must also reproduce the ordered joint distribution,
\begin{equation}
P(m_1,m_2)
=
f_A(m_1)f_B(m_2)+f_B(m_1)f_A(m_2).
\end{equation}
This condition fixes the assignment up to a global exchange of $A$ and $B$.

\paragraph{Case III.}
For an exchange-symmetric labeled distribution,
Eq.~\eqref{eq:ordering_case_III_joint} can be inverted by reflecting the ordered density across the equal-mass line:
\begin{equation}
    f_{\rm III}(m_{\rm A},m_{\rm B})
    =
    \begin{cases}
        \dfrac{1}{2}P(m_{\rm A},m_{\rm B}),
        & m_{\rm A}>m_{\rm B},\\[7pt]
        \dfrac{1}{2}P(m_{\rm B},m_{\rm A}),
        & m_{\rm B}>m_{\rm A}.
    \end{cases}
    \label{eq:case3_reconstruction}
\end{equation}
This uniquely determines the exchange-symmetric labeled distribution.

\paragraph{Case IV.}
For an arbitrary joint labeled distribution,
Eq.~\eqref{eq:ordering_case_IV_joint} determines only the exchange-symmetric combination
\begin{equation}
    f_+(m_{\rm A},m_{\rm B})
    =
    \frac{
        f(m_{\rm A},m_{\rm B})
        +
        f(m_{\rm B},m_{\rm A})
    }{2}.
\end{equation}
The exchange-antisymmetric part,
\begin{equation}
    f_-(m_{\rm A},m_{\rm B})
    =
    \frac{
        f(m_{\rm A},m_{\rm B})
        -
        f(m_{\rm B},m_{\rm A})
    }{2},
\end{equation}
does not affect the ordered distribution. Any allowed \(f_-\) can therefore be added without changing \(P(m_1,m_2)\), so the full labeled mass plane cannot be uniquely reconstructed in Case IV.

\section*{Supplemental Material III: High-mass population inference}
\label{app:population_inference}

\begin{figure*}[t]
    \centering
    \includegraphics[width=0.99\textwidth]{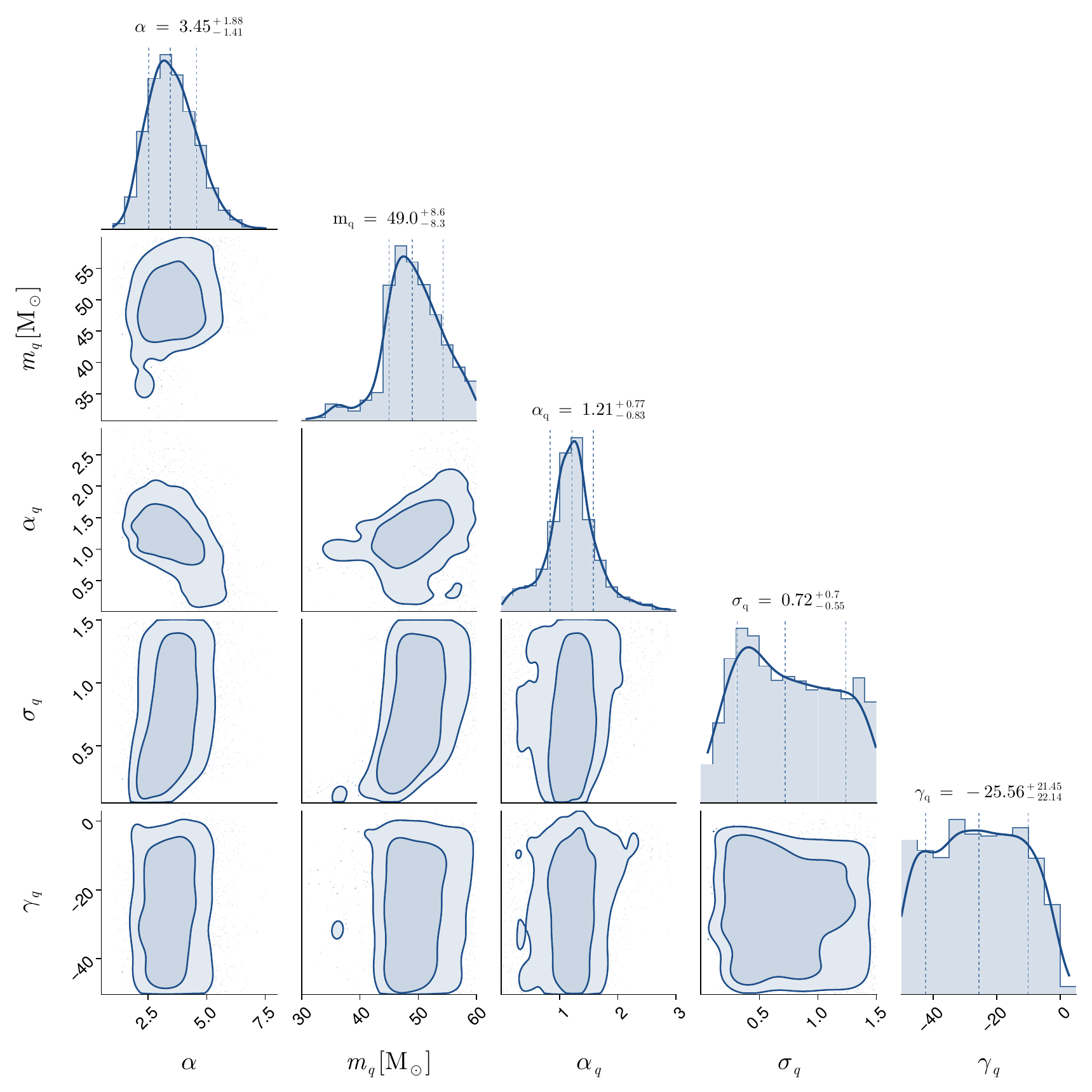}
    \caption{
        Joint posterior distributions for the high-mass population hyperparameters. The anticorrelation between \(\alpha_q\) and \(\alpha\) indicates that decreasing \(\alpha_q\), which tilts the ridge toward equal mass, is accompanied by a steeper primary-mass falloff. The positive correlation between \(\alpha_q\) and \(m_q\) compensates for changes in the ridge location: increasing \(\alpha_q\) shifts the ridge downward, whereas increasing \(m_q\) shifts it upward. Together, these correlations show how variations in \(\alpha\) and \(m_q\) compensate for departures from the constant-\(m_2\) scaling, \(\alpha_q=1\). Code used to produce this figure is available \href{https://github.com/SoumendraRoy/Mass_Ordering/blob/main/scripts/Plots_Main.ipynb}{here}.
    }
    \label{fig:high_mass_corner}
\end{figure*}

\begin{figure}[t]
    \centering
    \includegraphics[
        width=\columnwidth
    ]{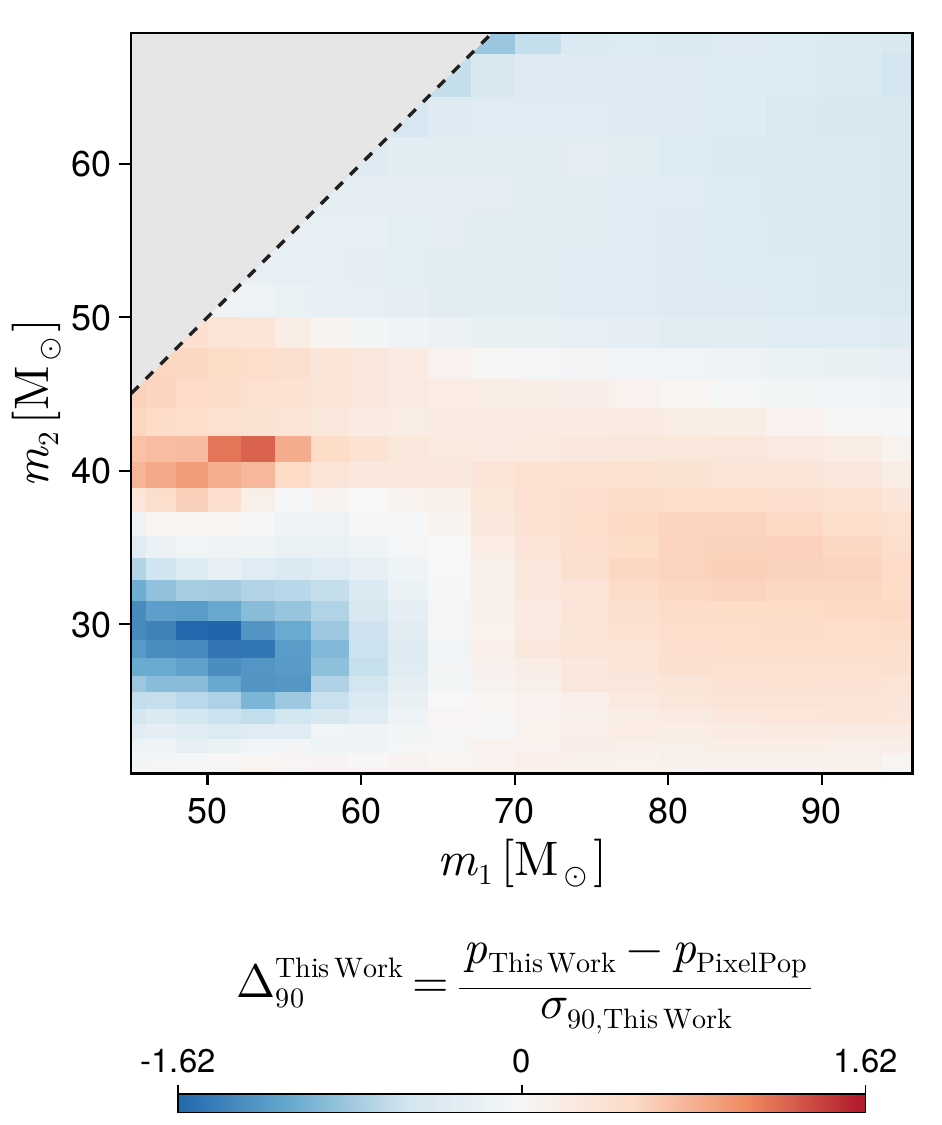}
    \caption{
        Comparison of the normalized mass distributions inferred with our parametric model and the \(m_1\)--\(m_2\) \texttt{PixelPop} analysis reported in the GWTC-5 population paper \citep{2026arXiv260527226T}. The color scale shows \(\Delta_{90}^{\rm This\,Work}=(p_{\rm This\,Work}-p_{\rm PixelPop})/\sigma_{90,\rm This\,Work}\), using the 90\% uncertainty of our model. The \texttt{PixelPop} median lies within this interval in 94.6\% of the common ordered grid cells, with a maximum deviation of \(1.6\,\sigma_{90}\approx2.6\sigma\). The code used to produce this figure is available \href{https://github.com/SoumendraRoy/Mass_Ordering/blob/main/scripts/Plots_Main.ipynb}{here}.
    }
    \label{fig:gwtc5_comparison}
\end{figure}

\begin{figure*}[t]
    \centering
    \includegraphics[
        width=\textwidth,
        keepaspectratio
    ]{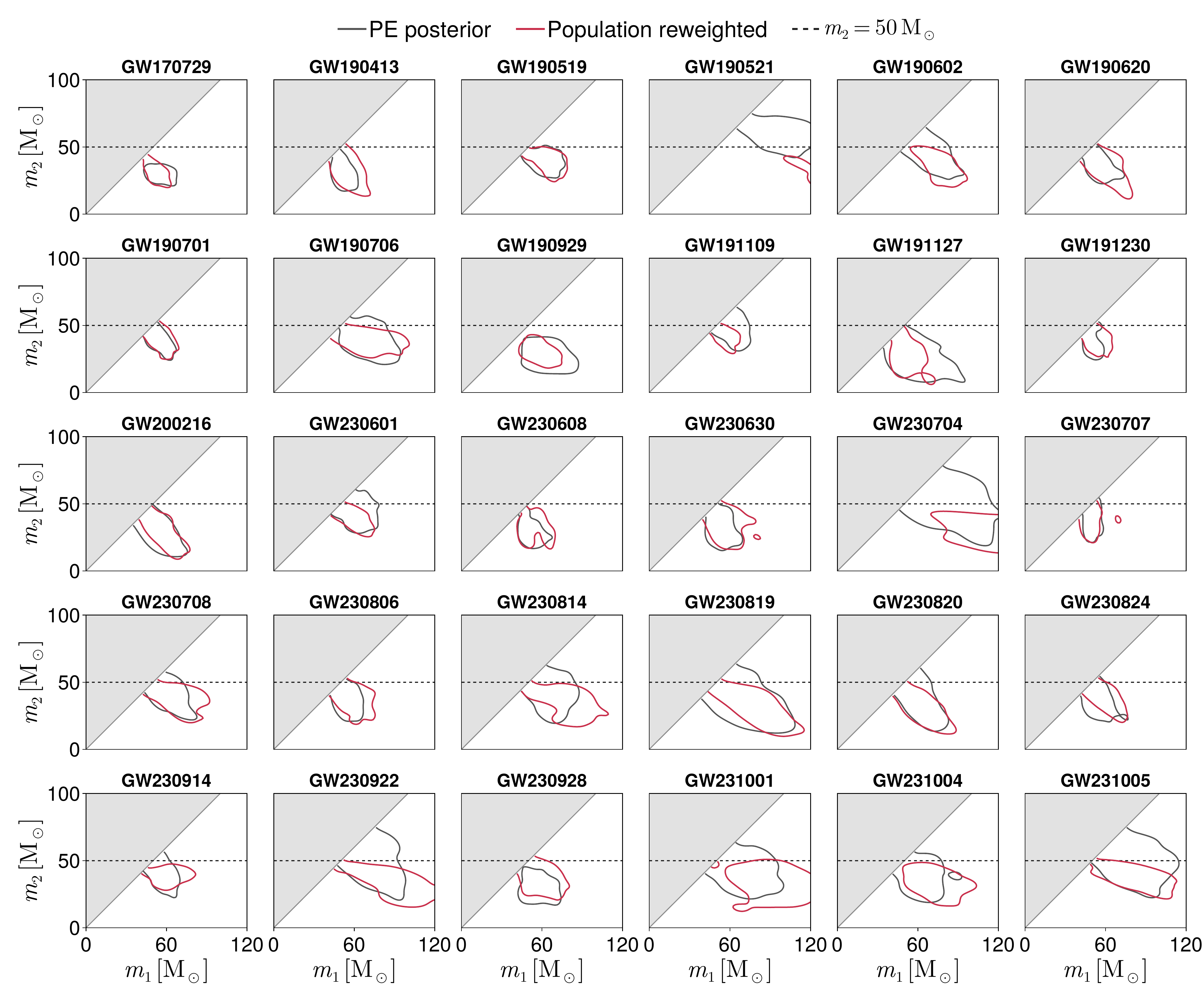}
    \caption{
        Event-level source-frame mass posteriors for the first 30 high-mass binary black hole mergers. Gray contours enclose \(90\%\) of the original parameter-estimation posterior, while red contours enclose the corresponding \(90\%\) population-reweighted credible region. The dashed line marks the characteristic secondary-mass scale, \(m_2=50\Msun\). The code used to produce this figure is available \href{https://github.com/SoumendraRoy/Mass_Ordering/blob/main/scripts/Plots_Main.ipynb}{here}.
    }
    \label{fig:event_reweighting_1}
\end{figure*}

\begin{figure*}[t]
    \centering
    \includegraphics[
        width=\textwidth,
        keepaspectratio
    ]{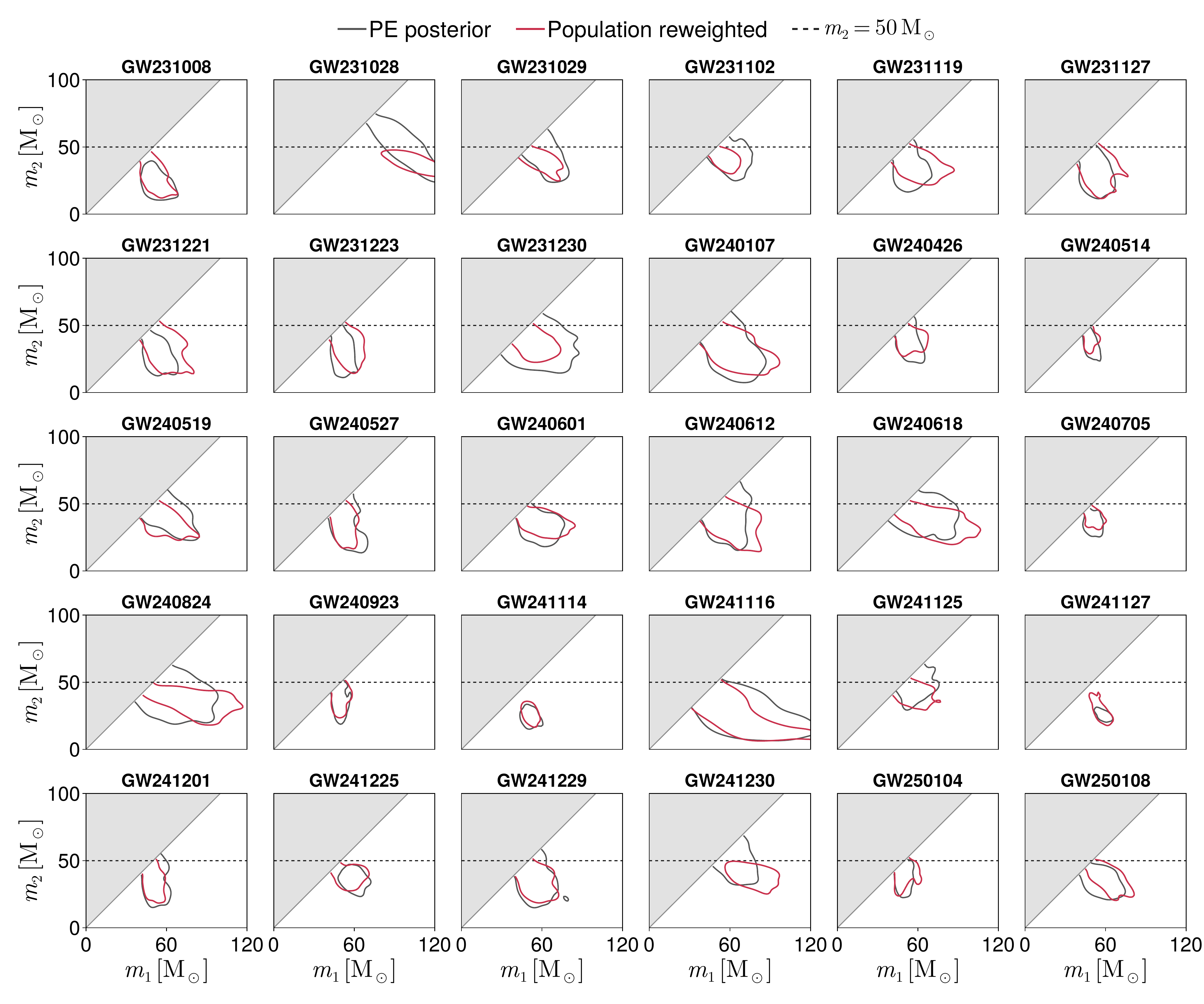}
    \caption{
        Same as Fig.~\ref{fig:event_reweighting_1}, but for the remaining 30 high-mass binary black hole mergers. Most events favor \(m_2\lesssim50\Msun\), while a smaller subset retains support at larger secondary masses and produces the high-\(m_2\) tail of the inferred population. The code used to produce this figure is available \href{https://github.com/SoumendraRoy/Mass_Ordering/blob/main/scripts/Plots_Main.ipynb}{here}.
    }
    \label{fig:event_reweighting_2}
\end{figure*}

\begin{figure}[t]
    \centering
    \includegraphics[
        width=\columnwidth
    ]{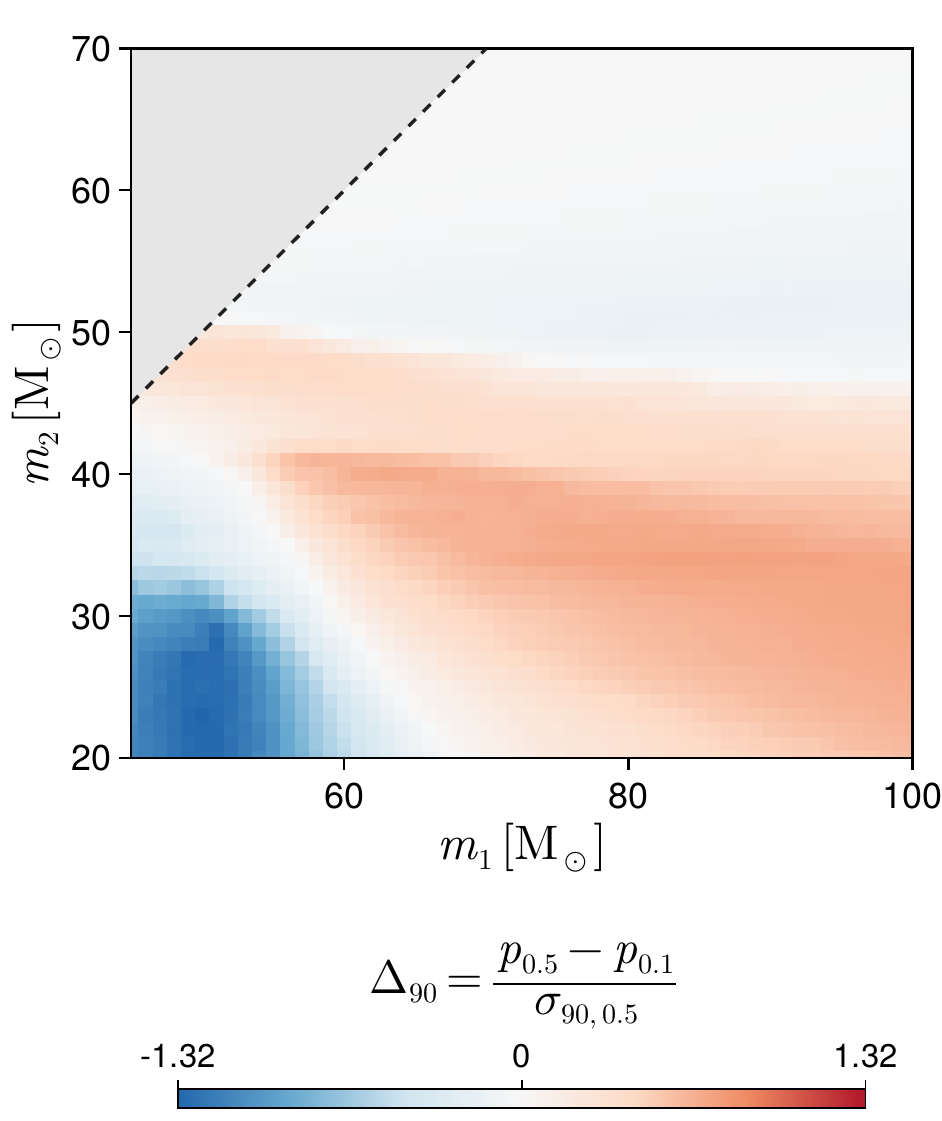}
    \caption{
        The normalized ordered-mass distribution for \(m_1>45\,\Msun\) is robust to variations in the posterior mass-selection threshold \(p_{\rm cut}\). The \(p_{\rm cut}=0.1\) median lies within the 90\% credible interval for \(p_{\rm cut}=0.5\) over most of the mass plane. The maximum variation is \(1.32\,\sigma_{90}\approx2.2\,\sigma\), where \(\sigma\) denotes the Gaussian standard deviation. The code used to produce this figure is available \href{https://github.com/SoumendraRoy/Mass_Ordering/blob/main/scripts/Plots_Main.ipynb}{here}.
        }
    \label{fig:pcut_robustness}
\end{figure}

\begin{figure*}[t]
    \centering
    \includegraphics[width=0.95\textwidth]{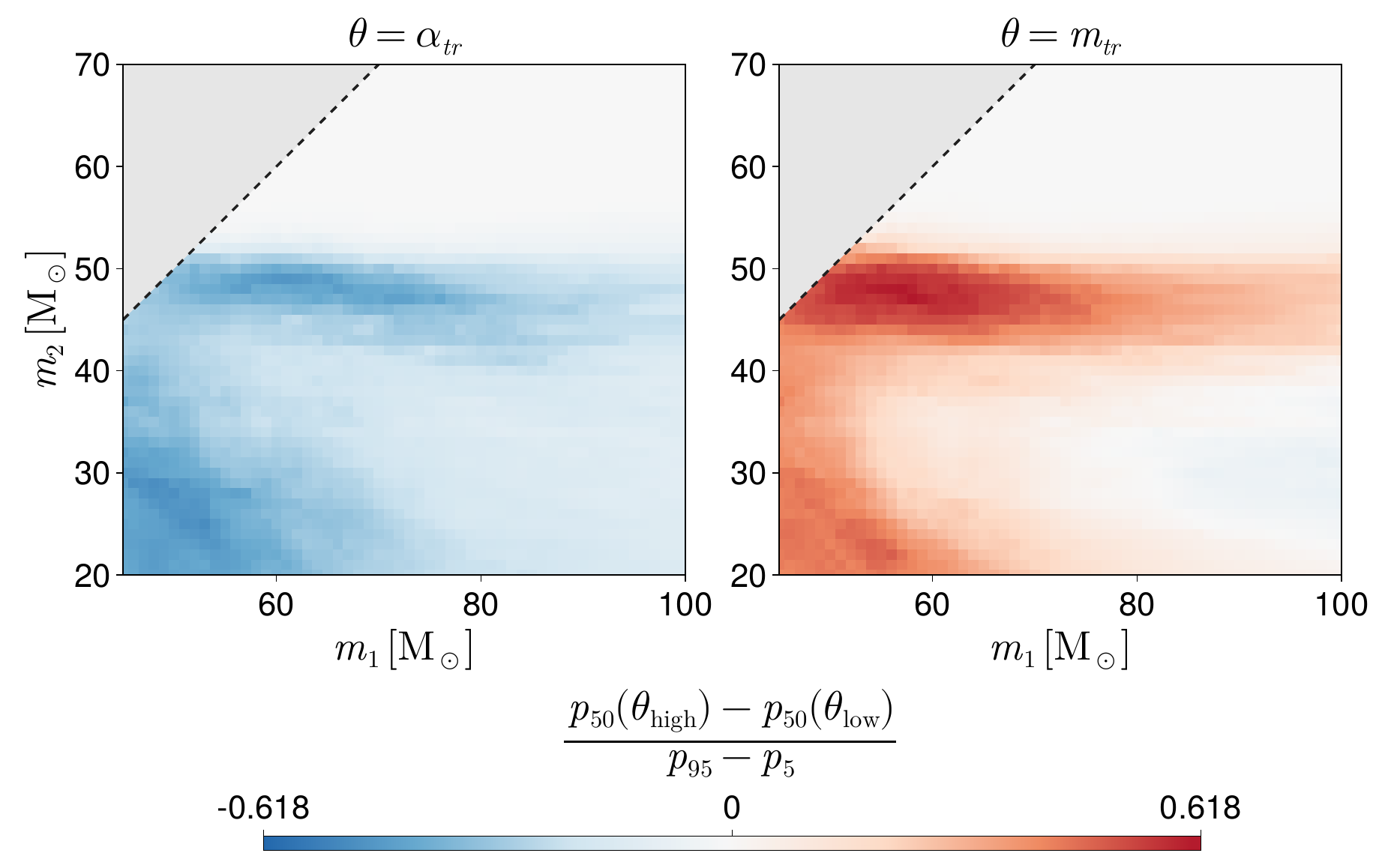}
    \caption{
        Sensitivity of the normalized ordered-mass distribution to the parameters governing the region below the mass cut. The left panel conditions on \(\theta=\alpha_{\rm tr}\), and the right on \(\theta=m_{\rm tr}\). At each mass pair, \(p_{50}(\theta_{\rm high})\) and \(p_{50}(\theta_{\rm low})\) are the median densities from draws in the highest and lowest 25\% of posterior values of \(\theta\), respectively. The denominator \(W_{90}=p_{95}-p_5\) is the pointwise 90\% interval width from all draws. The differences are generally small compared with this width and reach at most \(0.62\,W_{90}\). The inferred distribution above the cut is relatively insensitive to \(\alpha_{\rm tr}\) and \(m_{\rm tr}\), consistent with weak coupling across the mass cut. The code used to produce this figure is available \href{https://github.com/SoumendraRoy/Mass_Ordering/blob/main/scripts/Plots_Main.ipynb}{here}.
    }
    \label{fig:lower_mass_sensitivity}
\end{figure*}

Let \(\vec{\theta}\) denote the binary parameters and
\(\vec{\lambda}\) the population hyperparameters. We infer
\(\vec{\lambda}\) from a catalog of \(N_{\rm obs}\) events that pass two selection criteria, denoted by \({\rm det1}\) and \({\rm det2}\).

\paragraph{Hierarchical likelihood.}
The hierarchical likelihood is \citep{roy2025midthirtiescrisisdissectingproperties}
\begin{equation}
\begin{aligned}
    \mathcal{L}(\vec{d}\mid\vec{\lambda})
    &=
    e^{-N_{\rm det1,det2}(\vec{\lambda})}\\
    &\quad\times
    \prod_{i=1}^{N_{\rm obs}}
    \int d\vec{\theta}\,
    \frac{
        P_{\rm PE}(\vec{\theta}\mid d_i)
    }{
        \pi_{\rm PE}(\vec{\theta})
    }
    \frac{dN}{d\vec{\theta}}(\vec{\lambda}),
\end{aligned}
\label{eq:hierarchical_likelihood}
\end{equation}
where \(P_{\rm PE}(\vec{\theta}\mid d_i)\) is the parameter-estimation (PE) posterior for event \(i\), \(\pi_{\rm PE}(\vec{\theta})\) is the corresponding prior, and \(N_{\rm det1,det2}(\vec{\lambda})\) is the expected number of events that pass both selection criteria.

The first selection criterion is
\begin{equation}
    {\rm det1}:
    \qquad
    {\rm FAR}<1\,{\rm yr}^{-1}.
    \label{eq:det1}
\end{equation}
The second criterion requires sufficient posterior support in the mass region
\begin{equation}
    \Omega
    =
    \left\{
        (m_1,m_2)
        \,\middle|\,
        m_1>45\Msun,\;
        5\Msun<m_2<70\Msun
    \right\}.
    \label{eq:mass_domain}
\end{equation}
Specifically, we require
\begin{equation}
    {\rm det2}:
    \qquad
    P_{\rm PE}\bigl((m_1,m_2)\in\Omega\bigr)>p_{\rm cut},
    \qquad
    p_{\rm cut}=0.5.
    \label{eq:det2}
\end{equation}
To assess the robustness of our results to the choice of \(p_{\rm cut}\), Fig.~\ref{fig:pcut_robustness} compares the inferred distributions under variations of \(p_{\rm cut}\).

After applying this event-level criterion, we retain only the posterior samples inside \(\Omega\). The upper bound \(m_2<70\Msun\) excludes GW231123 \citep{2025ApJ...993L..25A} and restricts the analysis to the population associated with the high-mass ridge.

The expected number of selected events is
\begin{equation}
    N_{\rm det1,det2}(\vec{\lambda})
    =
    \int d\vec{\theta}\,
    P({\rm det1,det2} \mid \vec{\theta})
    \frac{dN}{d\vec{\theta}}(\vec{\lambda}),
    \label{eq:expected_detections}
\end{equation}
where \(P({\rm det1,det2} \mid \vec{\theta})\) is the probability that an event with parameters \(\vec{\theta}\) passes both criteria.

We evaluate the event integrals using discrete posterior samples:
\begin{equation}
\begin{aligned}
    \mathcal{L}(\vec{d}\mid\vec{\lambda})
    &\simeq
    e^{-N_{\rm det1,det2}(\vec{\lambda})}\\
    &\quad\times
    \prod_{i=1}^{N_{\rm obs}}
    \sum_{\vec{\theta}\sim
    P_{\rm PE}(\vec{\theta}\mid d_i)}
    \frac{
        dN/d\vec{\theta}(\vec{\lambda})
    }{
        \pi_{\rm PE}(\vec{\theta})
    }.
\end{aligned}
\label{eq:pe_sample_likelihood}
\end{equation}
The sum for each event contains only the posterior samples inside \(\Omega\).

We estimate \(N_{\rm det1,det2}\) using detected injections drawn from \(p_{\rm draw}(\vec{\theta})\):
\begin{equation}
    N_{\rm det1,det2}(\vec{\lambda})
    \simeq
    \frac{1}{N_{\rm draw}}
    \sum_{\vec{\theta}\sim
    P({\rm det1},{\rm det2},{\rm draw})}
    \frac{
        dN/d\vec{\theta}(\vec{\lambda})
    }{
        p_{\rm draw}(\vec{\theta})
    }.
    \label{eq:injection_sum}
\end{equation}
We apply \({\rm det1}\) directly to the injections \citep{2023PhRvD.108d3011E, essick2025compactbinarycoalescencesensitivity}. For \({\rm det2}\), we generate mock PE samples for each injection following the procedure described in Appendix C of \cite{Roy_2025}, and impose the same posterior-probability threshold used for the observed events. We ensure that the Monte Carlo estimates are converged by requiring adequate effective sample sizes for both the event-level posterior integrals and the selection-function integral \citep{Tiwari_2018, Pdet1-Farr, Pdet2-essick, 10.1093/mnras/stad2968, heinzel2025nottrustmontecarlo}.

Although \({\rm det2}\) targets systems with \(m_1>45\Msun\), the population model must extend below this boundary. The cut is applied to the measured posterior rather than the true source parameters, so measurement uncertainty can move systems into or out of the selected sample. We include a lower-mass branch in the population model \eqref{eq:broken_powerlaw_m1} to account for this effect. We examine how the hyperparameters of this lower-mass branch affect the inferred mass distribution in Fig.~\ref{fig:lower_mass_sensitivity}.

\paragraph{Population model.}
We model the source-frame distribution in primary mass \(m_1\), mass ratio \(q=m_2/m_1\), and redshift \(z\). We fix the spin distribution to the median inferred by the GWTC-5 population analysis \citep{2026arXiv260527226T}.

The merger-rate density is
\begin{equation}
    \frac{d^2\mathcal{R}}{dm_1\,dq}
    =
    \mathcal{R}_0\,
    p(m_1)\,
    p(q\mid m_1)\,
    \frac{\psi(z)}{\psi(0)},
    \label{eq:full_population_model}
\end{equation}
where \(\mathcal{R}_0\) is the merger-rate density at \(z=0\).

For \(\vec{\theta}=(m_1,q,z)\), the expected number density entering the hierarchical likelihood is
\begin{equation}
    \frac{dN}{d\vec{\theta}}
    =
    T_{\rm obs}\,
    \frac{d^2\mathcal{R}}{dm_1\,dq}\,
    \frac{dV_c}{dz}\,
    \frac{1}{1+z}.
    \label{eq:expected_number_density}
\end{equation}

We use a broken power law for the primary-mass distribution:
\begin{equation}
    p(m_1)
    =
    C_m
    \begin{cases}
        \left(\dfrac{m_1}{m_{\rm tr}}\right)^{-\alpha_{\rm tr}},
        & 5\Msun\leq m_1<m_{\rm tr},\\[7pt]
        \left(\dfrac{m_1}{m_{\rm tr}}\right)^{-\alpha},
        & m_{\rm tr}\leq m_1\leq300\Msun,
    \end{cases}
    \label{eq:broken_powerlaw_m1}
\end{equation}
where \(C_m\) normalizes the distribution.

For the mass-ratio distribution, define
\begin{equation}
    m_{1,q}=\max(m_1,m_{\rm tr}).
\end{equation}
We then use
\begin{equation}
\begin{aligned}
    p(q\mid m_1)
    &=
    \operatorname{SkewNormal}_{[5\Msun/m_{1,q},\,1]}
    \Biggl[
        q\,\Bigm|\\[-0.2em]
    &\qquad
        \left(\frac{m_q}{m_{1,q}}\right)^{\alpha_q},
        \sigma_q,\gamma_q
    \Biggr].
\end{aligned}
\label{eq:mass_ratio_model}
\end{equation}
For \(m_1<m_{\rm tr}\), setting \(m_{1,q}=m_{\rm tr}\) fixes the mass-ratio distribution to its value at the transition. In the limit \(\gamma_q\rightarrow-\infty\), Eq.~\eqref{eq:mass_ratio_model} asymptotes to
\begin{equation}
\begin{aligned}
    p(q\mid m_1)
    &\longrightarrow
    \operatorname{Normal}_{
        [\,5\Msun/m_{1,q},\,q_{\max}\,]
    }
    \Biggl[
        q\,\Bigm|\\[-0.3em]
    &\qquad
        \left(\frac{m_q}{m_{1,q}}\right)^{\alpha_q},
        \sigma_q
    \Biggr],
    \\
    \mathrm{with~}q_{\max}
    &=
    \min\!\left[
        \left(\frac{m_q}{m_{1,q}}\right)^{\alpha_q},
        1
    \right].
\end{aligned}
\label{eq:gamma_q_limit}
\end{equation}
When \(5\Msun/m_{1,q}<\mu_q(m_1)<1\), this limit produces a Gaussian sharply truncated at its location parameter, creating an empty gap between the maximum merger-rate ridge and the equal-mass line.

Finally, we model the redshift evolution as
\begin{equation}
    \psi(z)
    =
    \frac{
        (1+z)^\lambda
    }{
        1+
        \left[
            \dfrac{1+z}{1+z_p}
        \right]^\kappa
    }.
    \label{eq:madau_dickinson}
\end{equation}
We fix the cosmological parameters to the values used to construct the GWTC-5 source-frame parameter-estimation samples.

\paragraph{Hyperparameter correlations.}
Fig.~\ref{fig:high_mass_corner} shows two notable degeneracies that are not apparent in Fig. \ref{fig:q_model_summary}. First, \(m_q\) is positively correlated with \(\alpha_q\). At fixed \(m_1>m_q\) and fixed ridge location, Eq.~\eqref{eq:m2loc} implies
\begin{equation}
    \frac{d\ln m_q}{d\alpha_q}
    = \frac{1}{\alpha_q}
    \ln\left(\frac{m_1}{m_q}\right)>0,
\end{equation}
so an increase in \(\alpha_q\), which shifts the ridge toward lower \(m_2\), is compensated by an increase in \(m_q\). Second, \(\alpha_q\) is anticorrelated with \(\alpha\). Decreasing \(\alpha_q\) below \(1\) tilts the ridge toward equal mass, while the accompanying increase in \(\alpha\) suppresses its high-\(m_1\) portion. Conversely, increasing \(\alpha_q\) above \(1\) tilts the ridge toward lower \(m_2\), while the accompanying decrease in \(\alpha\) gives greater weight to its high-\(m_1\) portion. Together, these correlations limit how strongly departures of \(\alpha_q\) from unity alter the inferred merger-rate geometry: \(m_q\) anchors the characteristic \(m_2\) scale, while \(\alpha\) down weights the high-\(m_1\) portion of ridges tilted in either direction away from the constant-\(m_2\) track.

\textit{Comparison with PixelPop.}
In Fig.~\ref{fig:gwtc5_comparison}, we compare our inferred merger-rate density for \(m_1>45\Msun\) with that obtained from the \(m_1\)--\(m_2\) variant of the \texttt{PixelPop} analysis \citep{2025PhRvD.111f3043H,2025arXiv250620731A} reported in the GWTC-5 population paper \citep{2026arXiv260527226T}. As shown in the figure, the two inferred merger-rate densities are consistent within their 90\% credible intervals.

\textit{Do individual events support the high-mass ridge?}
Figs.~\ref{fig:event_reweighting_1} and \ref{fig:event_reweighting_2} compare the original and population-reweighted mass posteriors for the 60 high-mass events in GWTC-5 satisfying the selection criteria \eqref{eq:det1} and \eqref{eq:det2}. For most events, the two contours overlap substantially and favor \(m_2\lesssim50\Msun\), supporting a characteristic secondary-mass scale near \(50\Msun\) followed by a rapid decline at higher masses. This scale does not define a strict upper cutoff. GW190521, GW230704, GW230922, GW231005, GW231028, GW240612, GW240824, GW241125, and GW241230 retain the strongest posterior support above \(50\Msun\) and contribute to the high-\(m_2\) tail of the inferred population. GW231123 remains an outlier relative to the high-mass population inferred here and is therefore not shown in Figs.~\ref{fig:event_reweighting_1} and \ref{fig:event_reweighting_2}.

GW190521 \citep{GW190521_Abbott_2020} shows the largest displacement between the original and population-reweighted contours. Its broad parameter-estimation posterior extends into the \(m_2>50\Msun\) region, where the inferred population density declines rapidly. Population reweighting therefore favors its lower-\(m_2\) support and shifts the contour downward. Because the same catalog, including GW190521, determines the population model, this shift reflects population conditioning and does not by itself identify the event as a population outlier \citep{Fishbach2020, 2026ApJ...996L...4M}.

\textit{Population-model misspecification.}
Our population model assumes a single smooth high-mass distribution, with mass-independent width and skewness in \(p(q\mid m_1)\), separable mass and redshift evolution \citep[see, for example,][]{Karathanasis2023}, and a fixed spin distribution. Additional subpopulations or a more complicated pairing structure could modify the detailed width and slope of the ridge. However, the characteristic mass scale is unlikely to be an artifact of these assumptions: the event-level posteriors show considerably less support above \(m_2\simeq50\Msun\) than below it, and our inferred merger-rate density agrees within the 90\% credible interval with the \texttt{PixelPop} analysis, which adopts substantially different modeling choices. These independent checks support the presence and approximate location of the characteristic mass scale, together with the faster decline toward higher \(m_2\) than toward lower \(m_2\). Interpreting this decline as a hard upper cutoff, or using it to infer a maximum black-hole mass, remains population-model dependent because such inferences are controlled by the sparsely populated tail and its most extreme events \citep{Fishbach2020, 2026ApJ...996L...4M}.

\paragraph{Robustness to the mass selection.}
We select events whose PE posterior assigns more than \(p_{\rm cut}=0.5\) probability to the region \(\Omega\) defined in Eq.~\eqref{eq:mass_domain}:
\[
P_{\rm PE}\bigl((m_1,m_2)\in\Omega\bigr)>p_{\rm cut}.
\]
The cut therefore acts on the PE posterior, and we apply the same criterion when evaluating the selection function. Near the mass boundary, fewer events constrain the population, so a flexible model has wider posterior predictive intervals as it approaches the cut.

We check the inference in two ways. First, we relax the threshold to \(p_{\rm cut}=0.1\), admitting more events. Fig.~\ref{fig:pcut_robustness} compares the resulting normalized ordered-mass distribution with the fiducial \(p_{\rm cut}=0.5\) result. The \(p_{\rm cut}=0.1\) median lies within the fiducial pointwise 90\% credible interval over most of the plotted mass plane. The maximum difference is \(1.32\,\sigma_{90}\approx2.2\,\sigma\), where \(\sigma\) denotes the Gaussian standard deviation.

Second, the region below the cut is described by two hyperparameters, \(\alpha_{\rm tr}\) and \(m_{\rm tr}\). Inaccuracies in the mock PE used to evaluate the selection function could affect the inference above the cut if it depends strongly on these parameters. Fig.~\ref{fig:lower_mass_sensitivity} compares normalized median distributions from draws in the lowest and highest 25\% of values of each parameter. The differences reach at most \(0.62\) times the full pointwise 90\% interval width. Together, these checks show that the inferred distribution above the cut is stable under the tested change in \(p_{\rm cut}\) and depends only weakly on \(\alpha_{\rm tr}\) and \(m_{\rm tr}\).

\section*{Supplemental Material IV: Prior and Posterior Predictive Ordered-Mass Marginals}\label{app:posterior_vs_prior}
A prior-to-posterior predictive comparison can establish whether the data favor different high-mass slopes for the ordered mass marginals within the assumed population model. If the prior-predictive $m_1$ and $m_2$ marginals have comparable slopes, while the posterior-predictive $m_2$ marginal declines more steeply, the data favor this asymmetry relative to the prior prediction. However, this does not establish that the asymmetry is independent of the model. The prior-predictive density combines the population distributions from different hyperparameter values, weighted by their prior probabilities. The posterior-predictive density instead uses their posterior probabilities after conditioning on the data. Both therefore depend on the assumed functional form and hyperparameter priors.

For example, two models can yield the same prior-predictive density in binary-parameter space while allowing different individual population distributions. The same data can then produce different posterior predictions under the two models.

Moreover, a change from comparable prior-predictive ordered slopes to a steeper posterior-predictive $m_2$ decline does not imply that the physically labeled $m_B$ marginal declines more steeply than the $m_A$ marginal, or that $m_B$ has a cutoff or characteristic mass scale. Ordering acts on both predictions, but its effect on the marginal slopes depends on the labeled population distribution. The data can therefore favor a stronger difference between the ordered slopes without requiring a corresponding difference between the labeled slopes. The predictive comparison establishes a preference for ordered-mass structure within the assumed model; interpreting that structure in terms of physically labeled components requires examining the ordering map described in this work.

\section*{Supplemental Material V: Mass Scales and Ridges in the Labeled Population}\label{app:ridge_labels}
The GWTC-5 high-mass BBH merger population exhibits a ridge at a preferred secondary mass near \(m_2\simeq50\,\Msun\) that remains approximately constant as \(m_1\) increases. This requires a characteristic mass scale in the labeled population as well. Whether the labeled \(m_{\rm A}\)--\(m_{\rm B}\) plane contains a horizontal or vertical ridge, however, depends on the assumed form of the labeled distribution. We examine this distinction using the four pairing cases introduced in the main text.

\paragraph{Case I.}
The labeled population \(f\) can be reconstructed from the ordered population \(P\) as
\begin{equation}
f(m_{\rm A},m_{\rm B})=
\begin{cases}
\frac{1}{2}P(m_{\rm A},m_{\rm B}),
& m_{\rm A}\geq m_{\rm B},\\[3pt]
\frac{1}{2}P(m_{\rm B},m_{\rm A}),
& m_{\rm A}<m_{\rm B}.
\end{cases}
\end{equation}
Thus, a horizontal ridge in the ordered \(m_1\)--\(m_2\) plane implies both a horizontal ridge and its vertical reflection in the labeled \(m_{\rm A}\)--\(m_{\rm B}\) plane.

\paragraph{Case II.}
The labeled distribution is
\(f(m_{\rm A},m_{\rm B})=f_{\rm A}(m_{\rm A})f_{\rm B}(m_{\rm B})\),
and the ordered distribution, for \(m_1 \geq m_2\), is
\begin{equation}
P(m_1,m_2)
=
f_{\rm A}(m_1)f_{\rm B}(m_2)
+
f_{\rm B}(m_1)f_{\rm A}(m_2).
\end{equation}
An exactly horizontal interior ridge has its conditional maximum at the same secondary mass \(m_\ast\) throughout a primary-mass interval \(I_{m_1}\). Differentiating at fixed \(m_1 \in I_{m_1}\) gives
\begin{equation}
f_{\rm A}(m_1)f_{\rm B}'(m_\ast)
+
f_{\rm B}(m_1)f_{\rm A}'(m_\ast)=0.
\end{equation}
Assuming \(f_{\rm B}(m_1)>0\) along the ridge, this becomes
\begin{equation}
r(m_1)f_{\rm B}'(m_\ast)+f_{\rm A}'(m_\ast)=0,
\qquad
\mathrm{with}~r(m_1)=\frac{f_{\rm A}(m_1)}{f_{\rm B}(m_1)}.
\end{equation}
If \(r\) takes different values at \(x,y\in I_{m_1}\), subtracting their equations yields
\begin{equation}
[r(x)-r(y)]f_{\rm B}'(m_\ast)=0.
\end{equation}
A horizontal ridge therefore has two possibilities, each with a different labeled plane.

\textit{(1) Constant ratio over \(I_{m_1}\).}
For \(r(m_1)=r_0\),
\begin{equation}
P(m_1,m_2)
=
f_{\rm B}(m_1)
\bigl[r_0f_{\rm B}(m_2)+f_{\rm A}(m_2)\bigr].
\end{equation}
If the bracketed sum has its maximum at \(m_\ast\), the ordered ridge is horizontal. Neither labeled distribution needs to peak there: their nonzero slopes can cancel. Thus, the labeled plane need not contain a horizontal or vertical ridge at \(m_\ast\).

\textit{(2) Varying ratio over \(I_{m_1}\).}
In this case, the stationarity condition requires
\begin{equation}
f_{\rm A}'(m_\ast)=f_{\rm B}'(m_\ast)=0.
\end{equation}
For a local maximum, we additionally require
\begin{equation}
f_{\rm A}(m_1)f_{\rm B}''(m_\ast)
+
f_{\rm B}(m_1)f_{\rm A}''(m_\ast)<0
\end{equation}
throughout \(I_{m_1}\). Since the labeled conditional distributions are
\begin{equation}
p(m_{\rm B}\mid m_{\rm A})=f_{\rm B}(m_{\rm B}),
\qquad
p(m_{\rm A}\mid m_{\rm B})=f_{\rm A}(m_{\rm A}),
\end{equation}
negative curvature in \(f_{\rm B}\) produces a horizontal labeled ridge at \(m_{\rm B}=m_\ast\), while negative curvature in \(f_{\rm A}\) produces a vertical ridge at \(m_{\rm A}=m_\ast\). The ordered maximum requires at least one of these labeled ridges, but not necessarily both.

\paragraph{Case III.}
Exchange symmetry, \(f(m_{\rm A},m_{\rm B})=f(m_{\rm B},m_{\rm A})\), gives the same reconstruction as Case I:
\begin{equation}
f(m_{\rm A},m_{\rm B})=
\begin{cases}
\frac{1}{2}P(m_{\rm A},m_{\rm B}),
& m_{\rm A}\geq m_{\rm B},\\[3pt]
\frac{1}{2}P(m_{\rm B},m_{\rm A}),
& m_{\rm A}<m_{\rm B}.
\end{cases}
\end{equation}
Thus, a horizontal ridge at \(m_2=m_\ast\) produces a horizontal ridge at \(m_{\rm B}=m_\ast\) on the \(m_{\rm A}>m_{\rm B}\) branch and its vertical reflection at \(m_{\rm A}=m_\ast\) on the \(m_{\rm B}>m_{\rm A}\) branch. Exchange symmetry requires both ridges to be present in the labeled plane.

\paragraph{Case IV.}
This is the general case relevant to our GWTC-5 inference: the mass model in Eq.~\eqref{eq:mass_model} imposes neither independence nor exchange symmetry on the physically labeled population. The ordered distribution constrains only
\begin{equation}
P(m_1,m_2)=f(m_1,m_2)+f(m_2,m_1).
\end{equation}
Defining \(w(m_1,m_2)\) as the fraction assigned to binaries in which \(A\) is heavier, we can write
\begin{equation}
\begin{aligned}
f(m_1,m_2)&=w(m_1,m_2)P(m_1,m_2),\\
f(m_2,m_1)&=[1-w(m_1,m_2)]P(m_1,m_2),
\end{aligned}
\end{equation}
with \(0\leq w\leq1\). Every such choice preserves the ordered distribution.

For constant \(0<w<1\), the labeled plane contains both the horizontal ridge and its vertical reflection. The limiting choices \(w=1\) and \(w=0\) retain only the horizontal or vertical ridge, respectively, as shown in Fig.~\ref{fig:ordering-degeneracy}. These limits assume that the preferred companion carries the same physical label throughout the ridge: \(B\) for \(w=1\), or \(A\) for \(w=0\). Mass-dependent weights can shift the labeled maxima, so neither branch must retain an axis-parallel ridge.

The GWTC-5 feature therefore requires a characteristic mass scale in the labeled population, but not a horizontal or vertical labeled ridge. If the preferred companion carries the same physical label throughout the observed ridge, the labeled plane contains a horizontal strip near \(m_{\rm B}=m_\ast\) or a vertical strip near \(m_{\rm A}=m_\ast\).

To summarize, a horizontal track at \(m_2\simeq m_\ast\) in the ordered \(m_1\)--\(m_2\) plane requires both a horizontal track and its vertical reflection in the labeled \(m_A\)--\(m_B\) plane in Cases I and III. In Cases II and IV, neither labeled track is guaranteed. However, in all four cases, either \(m_A\) or \(m_B\) must lie preferentially near a characteristic mass \(m_\ast\) over a region of the \(m_A\)--\(m_B\) plane, so that \(m_2=\min(m_A,m_B)\) remains near \(m_\ast\) and produces the horizontal \(m_1\)--\(m_2\) track. Therefore, the horizontal track at \(m_2\approx50\,\Msun\) observed in the \(m_1\)--\(m_2\) plane of the high-mass BBH merger population requires one of the physically labeled component masses to lie preferentially near \(50\,\Msun\) over a range of parameter space.
\end{document}